\documentclass[epsfig,amssymb,twocolumn,usenatbib]{mn2e}
\newcommand{\vecg}{\mbox{\boldmath $g$} {}}
\newcommand{\vecq}{\mbox{\boldmath $q$} {}}
\newcommand{\vece}{\mbox{\boldmath $e$} {}}
\newcommand{\na}{\mbox{\boldmath $\nabla$}{}}

\usepackage{epsfig}
\usepackage{aas_macros,graphicx,times,multirow,amsmath,bbold}
\newif\ifAMStwofonts
\title[MOND vs dark matter in low-mass galaxies]
{Low-mass disc galaxies and the issue of stability: MOND vs dark matter}
\author[S\'anchez-Salcedo et al.]
{F.~J.~S\'anchez-Salcedo$^{1}$\thanks{E-mail:jsanchez@astro.unam.mx}, E.~Mart\'{\i}nez-G\'{o}mez$^{2}$, 
V.~M.~Aguirre-Torres$^{2}$ and 
\newauthor H.~M.~Hern\'andez-Toledo$^{1}$
\\
$^{1}$Instituto de Astronom\'{\i}a, Universidad Nacional Aut\'onoma
de M\'exico, Ciudad Universitaria, Apt.~Postal 70
264, \\
C.P. 04510, Mexico City, Mexico\\
$^{2}$Departamento de Estad\'{\i}stica, Instituto Tecnol\'ogico Aut\'onomo de M\'exico (ITAM), Rio Hondo 1,
C.P. 01080, Mexico City, Mexico}

\begin{document}
\date{Accepted xxxx Month xx. Received xxxx Month xx; in original form
2016 April 12}
\pagerange{\pageref{firstpage}--\pageref{lastpage}} \pubyear{2007}
\maketitle

\label{firstpage}
\begin{abstract}
We analyse the rotation curves and gravitational stability of a sample of six
bulgeless galaxies for which detailed images reveal no evidence for strong bars.
We explore two scenarios: Newtonian dark matter models and MOdified Newtonian 
Dynamics (MOND). 
By adjusting the stellar mass-to-light ratio, dark matter models can match simultaneously
both the rotation curve and bar-stability requirements in these galaxies. To be consistent
with stability constraints, in two of these galaxies, the stellar mass-to-light ratio is a 
factor of $\sim 1.5-2$ lower than the values suggested from galaxy colours.
In contrast, MOND fits to the rotation curves are poor in three galaxies, perhaps because
the gas tracer contains noncircular motions. The bar stability analysis provides a
new observational test to MOND.
We find that most of the galaxies under study require abnormally-high levels of random
stellar motions to be bar stable in MOND. In particular, for the only galaxy in the sample 
for which the line-of-sight stellar velocity dispersion has been measured (NGC 6503), 
the observed velocity dispersion is not consistent with MOND predictions
because it is far below the required value to guarantee bar stability. 
Precise measurements of mass-weighted velocity dispersions in (unbarred and
bulgeless) spiral galaxies are crucial to test the consistency of MOND.

\end{abstract}

%\email{jsanchez@astroscu.unam.mx}

%\slugcomment{To be submitted to }
%\email{jsanchez@astroscu.unam.mx}


\section{Introduction}
The rotation curves of spiral galaxies are very useful to measure the mass distribution in these
objects. The discrepancy between the dynamical mass and the luminous mass is
considered as a strong evidence of the existence of dark matter in these galaxies.
Alternatively, the mass discrepancy can be interpreted as a breakdown of the Newtonian 
law of gravity. In particular, MOdified Newtonian Dynamics (MOND), in which 
the gravitational force is larger than the standard Newtonian gravitational force at 
accelerations below a value $a_{0}\sim 1.2\times 10^{-8}$ cm s$^{-2}$,
can account for the shape and amplitude of the rotation curves of spiral galaxies amazingly well,
except for a handful of galaxies 
\citep[e.g.,][]{san02,bot02,mil07,san07,swa10,gen11,san13,ran14,rod14,bot15}
%[for a review]{bot02,mil07}
%(see Sanders \& McGaugh 2002 for a review;
%Bottema et al.~2002; Milgrom \& Sanders 2007; Sanders \& Noordermeer 2007;
%Swaters et al.~2010; Gentile et al.~2011; S\'anchez-Salcedo et al.~2013; Rodrigues et al.~2014; 
%Bottema \& Pesta\~na 2015).
In practice, the quality of MOND fits to the rotation curves are sensitive to uncertainties 
in the adopted parameters of the galaxy such as inclination, distance or the adopted 
form of the interpolating function $\mu(x)$. In addition, noncircular motions and warps
may also affect the observed rotation curve. Thus, it is important to investigate other
gravitational effects which can offer independent tests 
\citep[see][for a compilation of evidence that supports MOND]{san02,fam12}.
%(see Sanders \& McGaugh 2002
%and Famaey \& McGaugh 2012, for a compilation of evidence that supports MOND). 

In the dark matter models, the separate contributions to the rotation curve
from the baryonic components and dark matter are not easily disentangled 
because equally well-fitting models to the rotation curve can be obtained
with different stellar mass-to-light ratios. For unbarred galaxies, disc 
stability constraints may provide upper limits on the surface density of 
stellar discs or, equivalently, to the stellar mass-to-light ratio 
$\Upsilon_{\star}$ \citep{ath87,fuc98,fuc08,lis09,pug10,don15}.
%(Athanassoula et al.~1987; Fuchs \& von Linden 1998; Fuchs 2008;
%Lisker \& Fuchs 2009; Puglielli et al.~2010; D'Onghia 2015).

It is usually argued that unbarred galaxies with slowly rising
rotation curves at the centre are stabilized against bar instabilities by dark haloes, whereas
unbarred galaxies with circular velocities $\ga 150$ km s$^{-1}$ use to present a steep inner rise in the 
rotation curve and may be stabilized by their hard centres \citep[e.g.,][]{sel01}.
%(e.g., Sellwood \& Evans 2001). 
In the case of dwarf galaxies and low-surface brightness (LSB) galaxies, many studies led to the 
generally accepted picture that many 
of these galaxies are dominated by dark matter at almost all radii, especially those galaxies that have
gently rising rotation curves \citep[e.g.,][]{car88, per96, cot00,deb01,swa11}
%C\^{o}t\'e et al. 2000; de Blok et al. 2001; Swaters et al.~2011). 
For instance, \citet{swa11} considered a
sample of $18$ ealy-type dwarf galaxies and found that if the stellar mass-to-light
ratio in the $R$-band is assumed to be near unity, as predicted by stellar population synthesis, all
galaxies except one are dominated by dark matter.
The dominant dark matter halo and the low surface density could explain the rarity of bars in 
LSB galaxies \citep{mih97}. Still, some LSB galaxies may be massive enough
to become bar unstable \citep{may04}. Rotating haloes, triaxial haloes,
the dynamical state of cosmological haloes, the physics of gas and baryons,
and tidal interactions may lead to bar formation even in some dwarf and LSB galaxies
\citep{wei02,cur06,sah13,gho16}.
%(Weinberg \& Katz 2002; Curir et al. 2006; Saha \& Naab 2013; Ghosh et al. 2016). 
Due to all these factors, predictions on the fraction of barred galaxies as
a function of Hubble type are difficult.

In MOND, the rotation curve sets the mass-to-light ratio of the stellar disc and thus
there is less freedom to adjust the surface density of the disc, which is a crucial 
determinant in the stability analysis. 
Although it is simple to show from the basic tenets of MOND, that discs with
low internal accelerations (deep MOND regime) are more stable than purely Newtonian 
discs \citep{mil14}, the added stability is limited. Indeed, discs in the deep MOND
regime are more unstable in MOND than in the equivalent Newtonian galaxy, that is the 
same disc but embedded in a rigid halo to have the same circular velocity. 
Given the limited stability provided by MOND, the following questions arise:
does MOND predict the correct level of stability in galaxies with gently rising rotation curves
at their centres? What about the stability of galaxies that are not in the deep MOND regime?
%McGaugh \& de Blok (1998) and Famaey \& McGaugh (2012) 
\citet{mcg98} and \citet{fam12} argued that MOND naturally
provides the required level of disc self-gravity to explain the existence
of bars and spiral structure observed in some (red) LSB galaxies (although the spiral arms
in these galaxies are fragmentary, extremely faint and difficult to trace).

Stability studies in pressure-supported stellar systems can be also
used to discriminate between modified gravity theories and Newtonian gravity plus a 
dark matter halo. \citet{nip11} found that
MOND systems are more prone to radial-orbit instability than their equivalent Newtonian
systems. They suggested that stability constraints combined with their measured
velocity dispersion profiles in globular clusters (e.g., NGC 2419)
and dwarf spheroidal galaxies could provide crucial tests for MOND models.

In this work, we have selected six pure-disc galaxies having
high-quality rotation curves in their inner regions. We consider
separately the standard dark matter scenario and MOND. To be 
satisfactory, the models should be able to reproduce correctly 
the shape and amplitude of the rotation curves for reasonable
values of $\Upsilon_{\star}$, 
but also the level of gravitational stability of the disc in these galaxies.
Since the selected galaxies do not have a dense centre able to stabilize 
the stellar disc against bar-forming modes, high levels of instability would be very difficult to
reconcile with detailed images that reveal no evidence for the presence of 
strong bars. Moreover, if MOND is a good effective theory
(albeit not definitive) to describe the phenomenology of galaxies, the stability analysis
in the MOND framework may be used as a tool to make testable predictions. 

The paper is organised as follows. 
In Section \ref{sec:basis} we present the theoretical basis and the disc-stability criteria in 
Newtonian dynamics and in MOND.
In Section \ref{sec:sample} we briefly describe the properties of the
selected galaxies. Section \ref{sec:Newtonian_analysis} 
presents the fits to the rotation curves and the stability analysis under 
the standard dark matter scenario. In Section \ref{sec:MOND_analysis}, we perform the
same analysis but in the MOND framework.
Final comments and conclusions can be found in Section \ref{sec:conclusions}.
A statistical Appendix is provided to describe the Bayesian approach used
to draw inferences on the relevant parameters and functions of these parameters.

\section{Stability constraints: theoretical framework}
\label{sec:basis}
\subsection{Newtonian dynamics}
There is a solid embodiement to suggest that the multicomponent (cold gas, warm gas
and stars) Toomre 
stability parameter $Q$ lies in the range $1<Q<2.5$ within the optical radius of
disc galaxies. 
The Toomre parameter is required to be $>1$ to avoid a violent fragmentation 
instability but it is generally less than $2.5$ in order to allow the discs to have a level of 
self-gravity enough to develop 
spiral structures and self-regulated star formation  
\citep{sel84,ath86,bot93,bot03,fuc99,fuc01,fuc08,kho03,zas04,ros12,rom13,for14}.
%(Sellwood \& Carlberg 1984; Athanassoula \& Sellwood 1986; Bottema 1993, 2003; 
%Fuchs 1999, 2001, 2008; Khoperskov et al.~2003; Zasov et al.~2004; Roskar et al.
%2012; Romeo \& Falstad 2013; Forbes et al. 2014). 

\subsubsection{Spiral arms and the $X_{m}$-parameter}
\label{sec:spiralarms}

In cold discs (i.e. $1<Q<2.5$), swing amplification of perturbations
in the disc may lead to the grow of non-axisymmetric structures (Toomre 1981).
For a disc with surface density $\Sigma$ and epicyclic frequency $\kappa$, 
the wavelength that preferently grows, $\lambda_{\rm max}$, can be written 
as $\alpha \lambda_{\rm crit}$ where
\begin{equation}
\lambda_{\rm crit}=\frac{4\pi^{2} G\Sigma}{\kappa^{2}},
\end{equation}
and $\alpha$ is a dimensionless factor which depends on the local derivative
of the circular velocity. For instance, if the rotation curve is parameterized by a power-law,
$v_{c}\propto R^{\beta}$, then $\alpha\simeq 2$ when $\beta=0$ (flat rotation curve), 
and $\alpha\simeq 1.25$ when $\beta=0.4$ (Athanassoula 1984; Fuchs 2001; Mayer et al. 2001).  
For linearly rising rotation curves ($\beta\simeq 1$), the low rate of differential rotation suppresses 
the swing amplification. However, resonances may support global modes (Lynden-Bell 1979) and 
the spiral structure is difficult to predict.

The density wave theory predicts that the number of spiral arms 
or multiplicity $m$ is given by
\begin{equation}
m(R)=\frac{2\pi R}{\alpha \lambda_{\rm crit}}=\frac{R\kappa^{2}}{2\pi G\alpha \Sigma}.
\label{eq:m_Newton}
\end{equation}
It is convenient to define the $X_{m}$-parameter, which depends on $R$, as
\begin{equation}
X_{m}(R)=\frac{R\kappa^{2}}{2\pi G m\Sigma}.
\label{eq:Xm_Newton}
\end{equation}
The maximum amplification of the mode
with multiplicity $m$ will occur at those radii for which $X_{m}\simeq \alpha$
(Toomre 1981; Athanassoula 1984; 
Athanassoula et al.~1987; Fuchs 2001; D'Onghia 2015). 
According to Equation (\ref{eq:Xm_Newton}), discs of a fixed rotation curve 
should exhibit higher spiral-arm multiplicity (and a smaller amplitude of the
arms) when the disc mass is decreased.
Combining Eqs. (\ref{eq:m_Newton}) and (\ref{eq:Xm_Newton}), it is simple to show
that the predicted number of arms can be determined from $X_{2}$ through the
relation $m(R)=2\alpha^{-1} X_{2}$.

The spiral arm multiplicity has been used as a powerful tool 
to constrain the mass of the disc in spiral galaxies and, thereby, the stellar mass-to-light ratio 
\citep{ath87,fuc98,fuc99,fuc08,pug10,pug11,don15}.
%(Athanassoula et al.~1987; Fuchs \& von Linden 1998;
%Fuchs 1999, 2008;  Puglielli et al.~2010; Puglielli 2011; D'Onghia 2015).
The analysis above is valid under the assumption that the stellar discs
are dynamically cold (i.e. $Q\la 2.2-2.5$). In hot discs, however, the
amplification factor due to the swing amplification may be highly suppressed
\citep{too81,ath84,fuc01,gho14}.
%(Toomre 1981; Athanassoula 1984; Fuchs 2001; Ghosh \& Jog 2014). 
Still, galaxies may present occasional, weak spiral structures due to 
resonances \citep{gho14} or interactions with companions.

\subsubsection{Bar-instability criteria}
If galactic discs are cold and massive, they may generate a large bar
structure. As a diagnostic of the expected morphology of a certain galaxy, 
it would be very useful to have a recipe to predict when
galaxies develop $m=2$ spiral arms or form a bar. Although $X_{2}$
might be used  as a diagnostic for bar formation (e.g., Mihos et al. 1997),
it was derived in linear theory and strictly it cannot provide a criterion to 
distinguish bar stable to bar unstable discs. One has to use numerical
simulations to study collective global instabilities. 

Efstathiou, Lake \& Negroponte (1982, hereafter ENL) derived a stability criterion
for bar formation in isolated (bulgeless) galaxies; cold 
exponential discs with a Toomre parameter close to but larger than $1$ are bar stable if 
\begin{equation}
{\mathcal{R}}\equiv \frac{v_{\rm max}}{(GM_{d}/h_{R})^{1/2}}>1.1,
\end{equation}
where $v_{\rm max}$ is the maximum rotational velocity, $M_{d}$ is the mass of the disc
and $h_{R}$ its radial scale length.
The ENL criterion provides an approximate indicator for stability in cold galaxies
($1<Q<2$) lacking a dense core or bulge (Mayer \& Wadsley 2004; Yurin
\& Springel 2015; Barnes 2016). Once $v_{\rm max}$ and $h_{R}$ are measured,
this criterion sets an upper  value on $M_{d}$ and thereby on $\Upsilon_{\star}$, 
which we denote by $\Upsilon_{\star}^{\rm (ENL)}$.
It is important to note that the ENL criterion was derived for discs in rigid 
haloes\footnote{There do exist dark matter candidates that behave more rigidly than a 
collisionless fluid (e.g., scalar field dark matter, see Slepian \& Goodman 2012).}. If haloes 
are made of collisionless particles, the disc-halo interaction produces a destabilizing
influence (Athanassoula 2008; Sellwood 2016), which may lead to the 
formation of bars in galaxies that are bar stable by the ENL criterion. 
Thus, in bulgeless galaxies the $\Upsilon_{\star}^{\rm (ENL)}$ values should be 
considered as conservative upper limits, unless the level of random stellar 
motions are larger than usually assumed in these simulations. 

Puglielli et al. (2010, hereafter PWC) found that a strong bar 
is formed if two conditions are met. The first one is that the minimum value of $Q$
along the disc, denoted by $Q_{\rm min}$, is less than $2.2$.
The second condition is that $\left<X_{2}\right> < 2.7$,
where the brakets $\left<...\right>$ represent the average value 
over two radial scale-lengths (see also Puglielli 2011), but excluding the
central region $R<2h_{z}$ ($h_{z}$ the vertical scaleheight, which is assumed to
be one fifth of the radial scalelength), where the razor-thin disc approximation is not valid.
This condition imposes a maximum value for the stellar mass-to-light ratio, 
$\Upsilon_{\star}^{\rm (PWC)}$, for galaxies with no strong bars.

\subsection{MOND}
\label{sec:MONDformalisms}
We will consider MOND as a modification of gravity's law,
rather than a change in the inertia. We consider two popular `modified-gravity' 
formulations of MOND, which provide a field equation between the density
distribution of a system $\rho$ and its associated gravitational potential
$\Phi$. The first one is the non-linear extension
of the Poisson equation suggested by Bekenstein \& Milgrom (1984)
given by
\begin{equation}
\na\cdot \left[\mu\left(\frac{|\na\Phi|}{a_{0}}\right)\na\Phi\right] =4\pi G \rho,
\label{eq:MB84}
\end{equation}
where $a_{0}$ is a universal acceleration of the order of $10^{-8}$ cm s$^{-2}$, and 
$\mu(x)$ is some interpolating function with the property that $\mu(x)=x$ 
for $x\ll 1$ and $\mu(x)=1$ for $x\gg 1$. 
The interpolating function may have different forms. A family of $\mu$-functions that
satisfies the required asymptotes are
\begin{equation}
\mu(x)=\frac{x}{(1+x^{n})^{1/n}}
\label{eq:mufamily}
\end{equation}
\citep{mil08}. 
The case $n=2$ corresponds to the standard form
proposed by \citet{mil83}, whereas the case $n=1$ is the so-called ``simple'' $\mu$-function
suggested by \citet{fam05}.

The second one is the quasi-linear MOND (QUMOND) formulation developed by \citet{mil10}:
\begin{equation}
\nabla^{2}\Phi = \na \cdot \left[\nu \left(\frac{|\na\Phi_{N}|}{a_{0}}\right)\na\Phi_{N}\right],
\end{equation}
where $\Phi_{N}$ is the Newtonian potential, which satisfies the standard Poisson equation
$\nabla^{2} \Phi_{N}=4\pi G \rho$. 
The non-linear modified Poisson equation and QUMOND are equivalent in a spherical
distribution of mass if $\nu (y)=1/\mu(x)$, where $y=x\mu(x)$.

\subsubsection{The $X_{m}$ parameter in MOND}
Using the non-linear Poisson equation (Eq. \ref{eq:MB84}), \citet{mil89} found that the perturbations 
in MOND behave as in Newtonian dynamics but replacing the Newton constant $G$ 
by $\chi_{nl}G$, with $\chi_{nl}=1/[\mu^{+} (1+L^{+})^{1/2}]$,
where $L=d\ln\mu/d\ln x$, and $\mu^{+}$ and $L^{+}$ are the values of $\mu$ and $L$, 
just above the disc.
To obtain this result, he used the standard WKB approximation, where
the disc has negligible thickness and the perturbations are tighly wound. 
Therefore, the Toomre and $X_{m}$ parameters in the non-linear extension of the Poisson equation are
\begin{equation}
Q_{M}=\mu^{+}(1+L^{+})^{1/2} \frac{\sigma_{R} \kappa}{3.36 G \Sigma},
\label{eq:ToomreMOND}
\end{equation}
and
\begin{equation}
X_{m}(R)=\mu^{+}(1+L^{+})^{1/2} \frac{R \kappa^{2}}{2\pi G m\Sigma},
\label{eq:XmMOND}
\end{equation}
where $\sigma_{R}$ is the radial velocity dispersion.

The prefactor $\mu^{+}(1+L^{+})^{1/2}$ is in general $\leq 1$.
Consequently, MOND discs should exhibit a lower
spiral-arm multiplicity and a larger amplitude of the arms
than their equivalent Newtonian systems.
In the Newtonian regime, it holds that $L^{+}=0$ and $\mu^{+}=1$, and hence 
the prefactor is equal to $1$. In the deep MOND regime, $L^{+}\rightarrow 1$
and thus the prefactor is $\simeq \sqrt{2}\mu^{+}$. However, it is misleading to think that
galaxies in the deep MOND regime (i.e., galaxies with $\mu^{+}\ll 1$)
will have very small $X_{m}$-values. Indeed, $\kappa$ depends on $\mu$
because of the boost that MOND produces in the rotation velocity.
In Appendix \ref{sec:app_X2_Newton_vars}, we calculate $X_{m}$ in terms of Newtonian
quantities. We show that a MOND galaxy has the same $X_{m}$ as a Newtonian
galaxy with a nonresponsive dark halo whose mass is twice the mass of the disc, 
at most. In galaxies with internal accelerations $\ga a_{0}$, the added stability in MOND
is more limited; the same $X_{m}$ parameter can be achieved in a Newtonian disc by
adding an inert dark halo having a mass less than half the mass of the disc.

In general, the level of stability in MOND comes from the basic tenets of the MOND theory
\citep{mil14} and thus it is expected to depend only slightly on the formulation.
In Appendix \ref{sec:X2QUMOND}, we present the derivation of the $X_{m}$-parameter in the
QUMOND formulation. It turns out that the Toomre parameter and the $X_{m}$-parameter in 
QUMOND are slightly smaller than the values using the non-linear Poisson equation. In
the remainder of the paper, for definiteness, we adopt for $X_{2}$ the expression derived 
using the non-linear Poisson equation.

\subsubsection{Bar-instability criteria in MOND}
\label{sec:bar_ins_MOND}

According to Equation (\ref{eq:ToomreMOND}), $Q_{M}\leq Q_{N,eq}$,
where $Q_{N,eq}$ is the Toomre parameter in the equivalent Newtonian $+$ halo
system (that is, a halo is added to have the same circular velocity). Therefore, 
galaxies in MOND are expected to be more unstable, locally and globally, than their equivalent
Newtonian galaxies. \citet{bra99} tested numerically that
the growth rate of the $m=2$ Fourier mode is larger (or equal) in MOND discs
than it is in the equivalent Newtonian disc $+$ halo system. The formation and evolution of
a bar in MOND and in Newtonian dynamics were described in \citet{tir07}.
They found that the bars in MOND galaxies develop very soon as compared
to the Newtonian dark matter models. 
Since MOND discs are more unstable than their Newtonian dark matter analogs,
discs with ${\mathcal{R}}<1.1$ are also expected to be bar
unstable in MOND. In fact, the ENL criterion does not take
into account the enhanced self-gravity of the disc in MOND.

\begin{figure}
\centering\includegraphics[width=8.5cm,height=9.90cm]{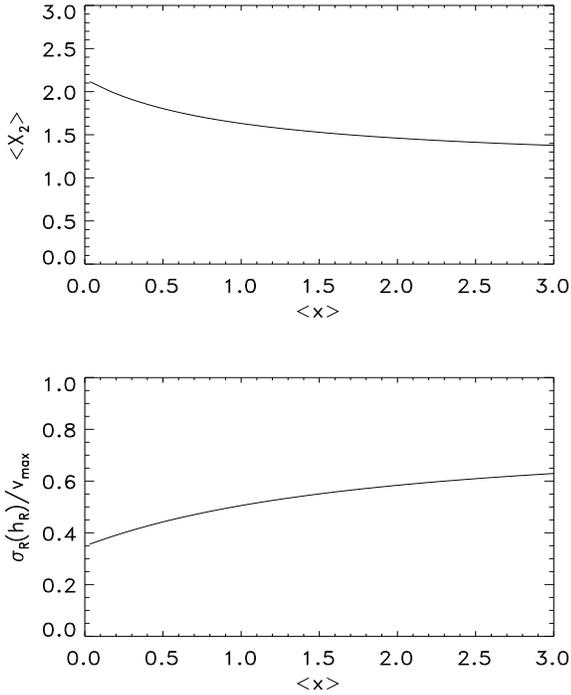}
 \caption{$\left<X_{2}\right>$ (top panel) and the ratio $\sigma_{R}/v_{\rm max}$
(with $\sigma_{R}$ evaluated at one radial scale length $h_{R}$) to have $Q_{M}=2.2$ 
(bottom panel), as a function of the
mean internal acceleration (in units of $a_{0}$). We have assumed a very thin 
exponential disc and used the simple interpolating function.}
\label{fig:X2_exp_disk}
\end{figure}

The PWC criterion for the formation of bars may be more restrictive than the ENL
criterion. To illustrate this, consider a pure-stellar disc having a negligible thickness 
and a surface density following an exponential profile.
Figure \ref{fig:X2_exp_disk} shows $\left<X_{2}\right>$ as a function of 
$\left<x\right>$ (where $x=|\na\Phi|/a_{0}$ was taken at the midplane of the disc)  for the 
simple $\mu$-function. Bare Newtonian discs have
$\left<x\right> \gg 1$, $\left<X_{2}\right>=1.15$, and are very unstable to bars.
When we move from bare Newtonian discs to MOND discs, 
$\left<X_{2}\right>$ increases but only slightly; $\left<X_{2}\right>$ is not
larger than $2.1$ even for galaxies in the deep MOND regime
($\left<x\right> \ll 1$). 
Indeed, it is not possible in MOND to satisfy the 
condition $\left<X_{2}\right> > 2.7$. This is a consequence of the limited
stability of MOND discs already discussed in \citet{mil89} and \citet{bra99}.
In other words, if galaxies have cold discs and do
not contain hard cores, disc self-gravity in MOND is strong enough to form bars, 
no matter how low the surface density of the disc is. In order for galaxies
with $\left<X_{2}\right> \leq 2$, as predicted in the MOND framework, 
to be bar stable, they need to be dynamically hot.  This may be accomplished
if the Toomre parameter is above some threshold value  $\sim 2.2$ \citep[e.g.,][]{ath86}.
The condition $Q_{M}=2.2$ provides a lower limit to the radial velocity dispersion $\sigma_{R}$.
Figure \ref{fig:X2_exp_disk} also shows the ratio between the radial velocity dispersion at one
scale length ($h_{R}$) and the maximum rotational velocity in order to match the condition
$Q_{M}=2.2$ in a thin exponential disc. We see that discs with $\left<x\right> \ga 1$ require 
$\sigma_{R} (h_{R})>0.5 v_{\rm max}$ to be bar stable.
Some questions arise: 
are unbarred galaxies (having no central hard cores) stable against bars
because the stellar discs have high velocity dispersions?
If the stellar discs were dynamically
heated by a bar which was finally dissolved, why do not all the galaxies have a 
prominent bulge?
The other less appealing possibility is that unbarred galaxies contain a 
stabilizing (unseen) spheroidal mass component, or a nonresponsive (unseen) 
thick disc, even in MOND.
In order to shed light on these issues and to illustrate how unbarred galaxies can be
used to test modified gravity theories, we have analysed a sample of bulgeless 
galaxies showing no strong evidence for bars.

 \begin{table*}
\begin{minipage}{140mm}
\caption[]{Comparison of the relevant parameters of the selected galaxies and references}
\vspace{0.01cm}

\begin{tabular}{c c c c c c c c}\hline
{Name} & {$D$} & {$M_{B}$} & {$\mu_{0}$} & {$i$} &  {$2h_{R}$} & References & References and  \\
{}&(Mpc) &  &  (mag arcsec$^{-2}$) &  (deg) & (kpc) &  for parameters & 
databases for Figs \ref{fig:images1}-\ref{fig:images2}  \\
\hline
NGC 3621 & $6.64 \pm 0.50$ &  $-20.05$ &  $18.2$ & $66^{\circ}$ &  $3.3$  & F01, dB08, G11 & SING \\
%\hline
NGC 4605 & 4.26$\pm$ 0.64 & -17.7 &  21.0  &$71.5^{\circ}$ & $1.36$ & S05 & SDSS-DR7, D09\\
NGC 5949 & $14.0\pm 2.4$ & -18.2 &  21.2 & $64.6^{\circ}$ &  $3.0$ & S05 & SDSS-DR7, D09\\
NGC 5963 & $13\pm 3$ & -17.8 &  20.1 & $48.4^{\circ}$ &  $1.36$   & S05 & SDSS-DR7, S10, J04\\
NGC 6503 & $5.2 \pm 1$ &  -17.7 &  19.2 & $74^{\circ}$ &  $2.8$  & K97, M99, B87 & L07, S10, D09\\
NGC 6689 & 11 &  -17.6 &  $20.8$ & $76.0^{\circ}$ &  $2.3$ &S05 & T05, SEIP, J04 \\

\hline
\end{tabular}
\medskip\\
Column (1): Name of the galaxy. Column (2): Distance. Column (3):
Absolute $B$ magnitude. Column (4): Extrapolated central surface brightness 
in the $B$-band, except for NGC 6689, which is in the $V$-band. 
Column (5): Photometric inclination.
Column (6): $2h_{R}$ is the distance at which the stellar surface density drops
a factor of $\exp(2)$ from its central value.
Column (7): References for the parameters
(F01: Freedman et al. 2001; dB08: de Blok et al. 2008; G11: Gentile et al.~2011;
S05: Simon et al.~2005; K97: Karanchentsev
\& Sharina 1997; M99: Makarova 1999; B87: Begeman 1987). Column (8): References
of the optical, infrared and continuum-subtracted H$\alpha$ images (SING: The
Spitzer Infrared Nearby Galaxies Survey, 2007, Fitfth Enhanced Data Release,
see Kennicutt et al. 2003 for a complete description;
SDSS-DR7: Sloan Digital Sky Survey Data Release 7, see Abazajian et al. 2009;
D09: Dale et al. 2009; S10: Sheth et
al. 2010; J04: James et al. 2004; L07: Lira et al. 2007; T05: Taylor et al. 2005; SEIP: 
The Spitzer Enhancing Imaging Products Database [http://irsa.ipac.caltech.edu/data/SPITZER/Enhanced/SEIP/]).
\label{table:basic_parameters}
\end{minipage}
\end{table*}

\section{The sample of galaxies: Photometric and kinematical data}
\label{sec:sample}
In the last decades, a great effort has been made to obtain
high-quality rotation curves in the central regions of LSB 
galaxies and dwarf irregular galaxies to test the
predictions of the standard cold dark matter model and the possible role
of the baryonic feedback in shaping the central density profiles of dark haloes. 
Here, we analyse the dynamics of one Sd galaxy and five low-luminosity Sc galaxies, 
having high-resolution rotation curves in the inner regions: 
NGC  3621, 4605, 5949, 5963, 6503 and 6689. 
All the galaxies are nearby ($D<15$ Mpc).
Table \ref{table:basic_parameters} summarizes the basic parameters of these galaxies.
The photometric inclinations of these galaxies vary between $48.4^{\circ}$ for NGC 5963 
to $76.0^{\circ}$ for NGC 6689.

\begin{figure*}
\centering\includegraphics[width=14.5cm,height=13.0cm]{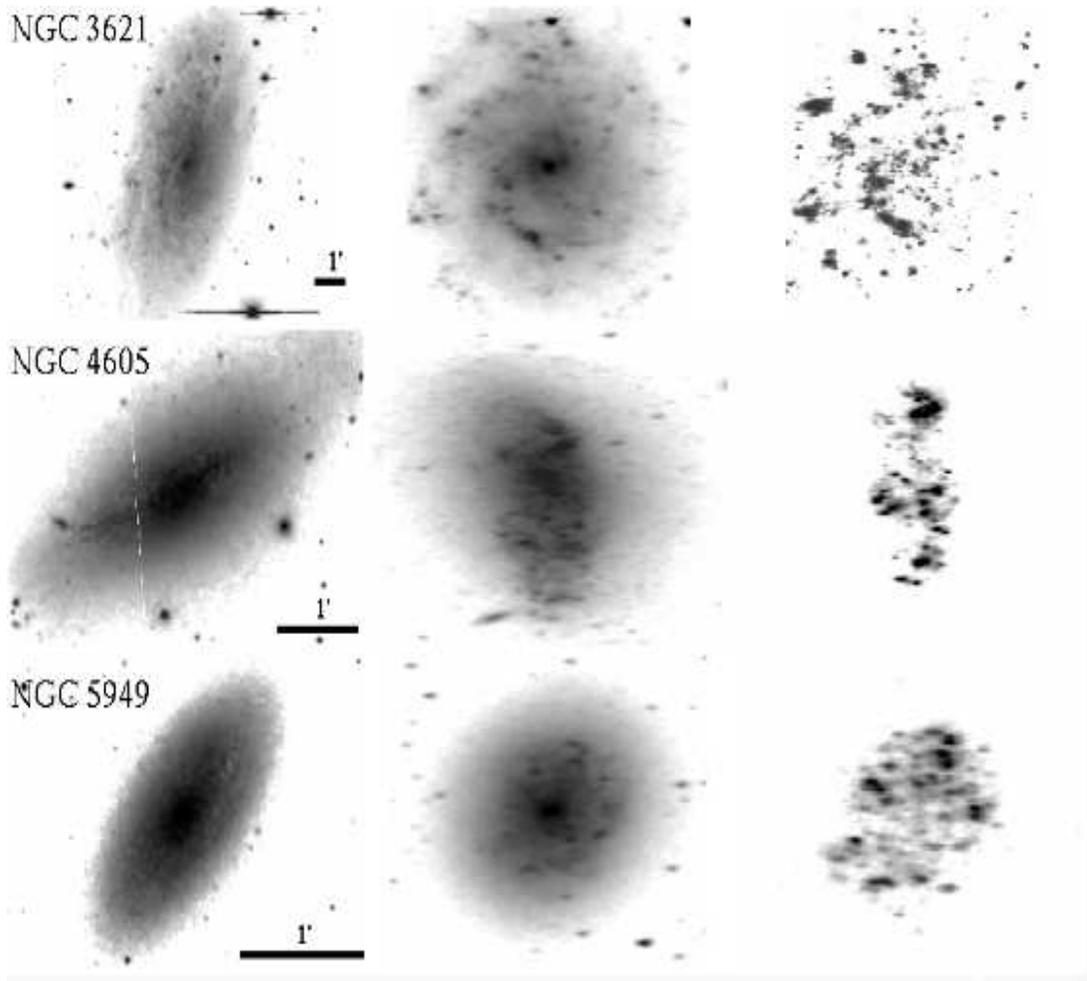}
 \caption{
Morphology of NGC 3621, 4605 and 5949 in different bands.
Left panels: Three-band (BVR/gri) optical images oriented according to the astronomical 
 convention (North up and East left). Middle panels: Deprojected Spitzer IRAC $3.6\mu$m images. 
Right panels: Deprojected 
 continuum-subtracted H$\alpha$ images. All images are displayed in logarithmic gray scale.
 See Table \ref{table:basic_parameters} for the relevant references
and databases used.}
\label{fig:images1}
\end{figure*}

\begin{figure*}
\centering\includegraphics[width=14.5cm,height=13.0cm]{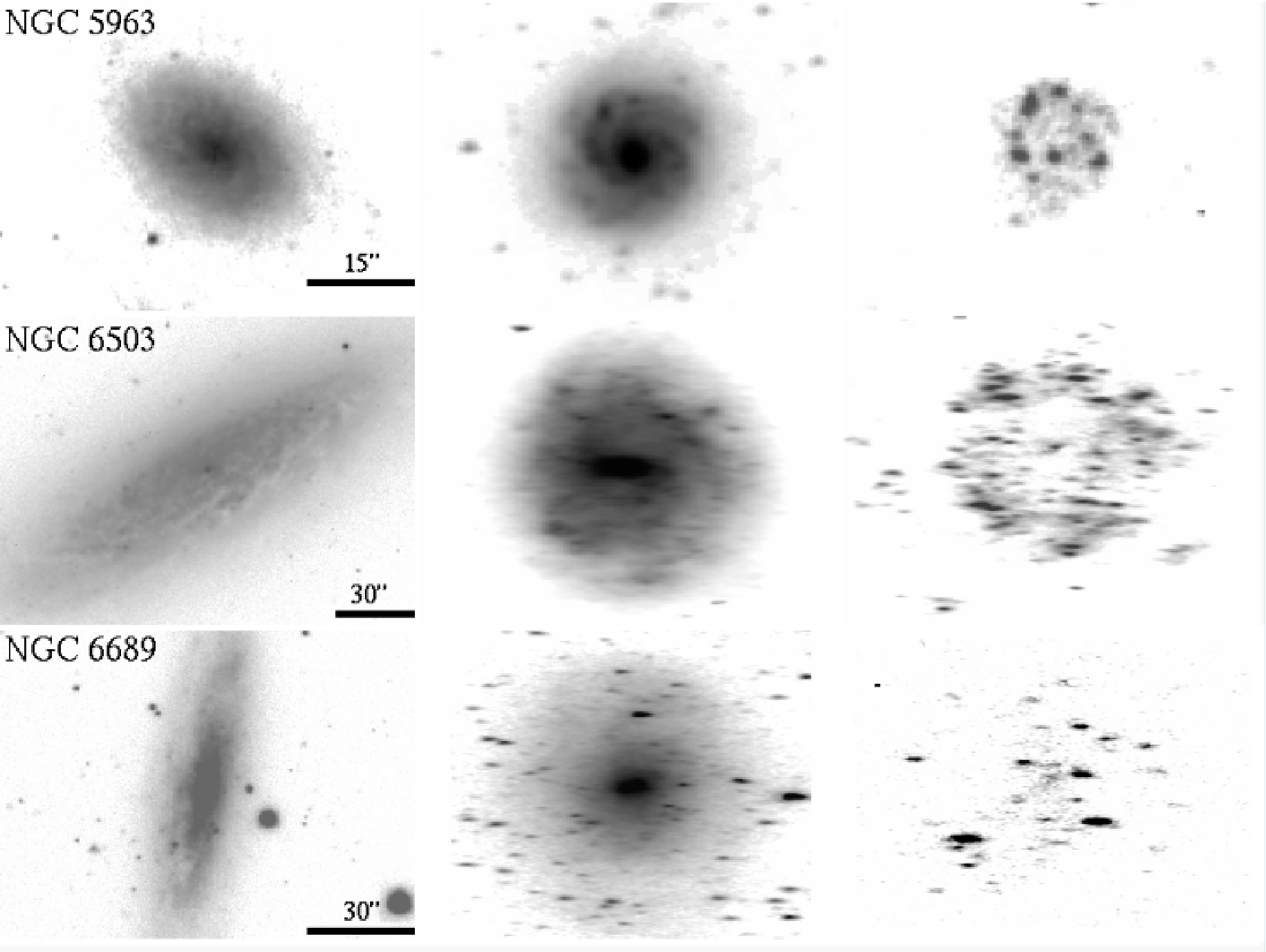}
 \caption{Same as Figure \ref{fig:images1} but for NGC 5963, 6503 and 6689.
See Table \ref{table:basic_parameters} for the relevant references
and databases used.}
\label{fig:images2}
\end{figure*}

Combined three-band optical images, Spitzer IRAC images at $3.6$ microns and
subtracted-continuum H$\alpha$ images are displayed in Figures \ref{fig:images1} 
and \ref{fig:images2}. In order to assess the morphology
of these galaxies, free from the effect of projection, the $3.6\mu$m and H$\alpha$ images
were deprojected.
To deproject each galaxy, flux-conserving stretching perpendicular to the line of nodes was carried out by using IRAF routines and by assuming that the outer disc should be intrinsically circular. 
Major axis position angles (North Eastwards), axis ratios (the major-to-minor axis ratio
up to the $25$ mag arcsec$^{-2}$ isophote in the $B$-band) and inclinations between the line of sight 
and the polar axis were all taken from HyperLeda database \citep{mak14}. 
 Although some of these galaxies were classified as barred (SBc) galaxies in 
some catalogs, there is no strong evidence for bars except in NGC 6503, which
 contains a faint end-on bar \citep{kuz12}.
NGC 3621 is representative of a pure-disc galaxy, with multiple spiral arms.
It is clear from the $3.6\mu$m image that NGC 4605 is not axisymmetrical. 
The deprojected H$\alpha$ image of NGC 4605 shows a bisymmetric morphology
and perhaps a central elongated structure. However, due to the relatively high inclination 
of NGC 4605, the resultant bar-like morphology in the deprojected H$\alpha$ image could
be somewhat spurious. Indeed, \citet[][hereafter S05]{sim05} did not find any 
strong evidence for a bar in NGC 4605.
The $3.6\mu$m and H$\alpha$ images show multiple spiral arms in NGC 5963. 
This galaxy also shows an elongated structure about $0.25$ kpc across. However,
it is probably a pseudobulge, not a bar (S05).
NGC 5949 has a more flocculent morphology, with less delineated
spiral arms. The high inclination of NGC 6689 precludes determining details
of its structure, though a deprojection shows signs of spiral-like structure both 
in the $3.6\mu$m and H$\alpha$ images. There is no kinematic evidence for a bar (see below).

%(Simon et al.~2005; hereafter S05).

To carry out the mass models, the photometric and kinematical data were 
taken from \citet[][hereafter dB08]{deb08}
for NGC 3621, from SO5 for NGC 4605, 5949, 5963 and 6689 and, from \citet{kuz12})
for NGC 6503.
S05 applied multicolour imaging to measure the isophotal parameters of the galaxies.
dB08 used the observed $3.6\mu$m surface brightness profile and
Kuzio de Naray et al. (2012) used the $K_{s}$-band. These bands are considered good
tracers of stellar mass.

NGC 3621 is a galaxy included in the H\,{\sc i} Nearby Galaxy Survey (THINGS)
described in \citet{wal08}, which consists of high-resolution H\,{\sc i}
observations of a sample of $34$ nearby galaxies. For NGC 5949 and NGC 6689, S05 
used H$\alpha$ two-dimensional velocity fields to derive the rotation curves. 
For NGC 4605, 5963 and 6503, H$\alpha$ and CO observations were combined
(S05; Kuzio de Naray et al.~2012). 
The kinematic and photometric values of the PA, centre and inclination angle agree within
the uncertainties.
Interestingly, S05 used a technique to incorporate the uncertainties of the PA,
inclination and centre position, obtaining more realistic rotation curve error bars than just
propagating only the uncertainties in the measure of the velocities from each spectrum.
\citet{spa08} also measured the H$\alpha$ rotation curve
of NGC 5949 and obtained an almost identical rotation curve as S05.

The tilted-ring models allowed to extract the rotational, radial
and systemic velocities as a function of radius from the velocity
fields. This is done by decomposing the velocity field in Fourier
components along the azimuthal angle $\theta$. In a purely 
axisymmetric galaxy, only the components $\cos \theta$ (rotation) and
$\sin \theta$ (radial velocity) are nonzero. The higher order terms
(e.g. $\cos 2\theta$, $\sin 2\theta$ and so on), if present, are
consequence of non-axisymmetric perturbations in the velocity field,
such a bars or spiral arms. For the S05 galaxies, S05 found that the 
amplitude of the higher order components is not large enough to affect 
the derived rotation curve. Moreover, this also suggests that bars, 
if present, are too weak to significantly disturb the kinematics of the gas.

For the galaxies studied, the non-circular motions are dominated by the radial component. 
Still, the radial motions are so small for NGC 3621, 5949, 6503 and 6689 that can be described
by rotation only. In the case of NGC 6689, no deviations from circular motions
were detected. This implies that if it has a bar, it should be very faint.
The radial motions are moderate in NGC 5963. NGC 4605 is the only galaxy in
the sample that deviates significantly from axisymmetry (S05). The radial motions
were interpreted as the result of the gas being in elliptical orbits due to the 
influence of a triaxial dark matter halo (S05).

The observed rotation curves for these galaxies are shown in Figure \ref{fig:RCs_iso}.
NGC 4605, 5949 and 6689 exhibit gently rising rotation curves. In the interval plotted,
the maximum circular velocities  lie between $100$ km s$^{-1}$ and
$140$ km s$^{-1}$. 
In the case of NGC 4605, \citet{rub80} already measured the H$\alpha$ rotation velocity.
Their observations extend a bit further, finding a rotational velocity of
$100$ km s$^{-1}$ at $3.3$ kpc in radius. For the galaxies NGC 3621, 5963 and 6503,
the H\,{\sc i} rotation curves extend up to $23$ kpc, $10.5$ kpc and $12.2$ kpc, respectively
(Bosma et al.~1988; dB08; Greisen et al.~2009; Kuzio de Naray et al.~2012). 
These studies show a flat rotation velocity of $150$ km s$^{-1}$ in NGC 3621,
$130$ km s$^{-1}$ in NGC 5963 and $116$ km s$^{-1}$ in the case of NGC 6503.

\section{Newtonian analysis}
\label{sec:Newtonian_analysis}
\subsection{Rotation curves and dark matter content}
\label{sec:RCs_DM}
Detailed Newtonian mass models for NGC 3621 can be found in dB08 \citep[see also][]{ran14}.
Mass models for NGC 4605, 5949, 5963 and 6689 were presented
in S05. On the other hand, the galaxy NGC 6503 has been extensively studied in the
literature because the observed velocity dispersion profile imposes additional 
constraints to the surface density of the disc \citep{bot89,bot97,fuc99,pug10}.
%(Bottema 1989; Bottema \& Gerritsen 1997; Fuchs 1999; Puglielli et al. 2010). 
The reader is referred to those papers for the details 
on the different dark matter mass models. Here we provide the rotation curves in dark matter 
models because they are useful as the basis of reference in making comparisons.

The light profiles were not decomposed into a disc and a bulge
component because all the galaxies are bulgeless except NGC 6503, which contains a small
pseudobulge. In this galaxy, the contribution of the 
pseudobulge to the total galaxy light is only $5$ percent \citep{kuz12}
and its contribution to the circular velocity is appreciable only within a radius of $200$ pc. Thus,
we can savely neglect the bulge in the mass budget. 

Following S05, the surface brightness corrected for inclination in the 
$R$-band is used as the reference band for NGC 4605, 5949 and 5963, 
and the $r'$-band for NGC 6689. For NGC 3621 and NGC 6503, we will use the $3.6\mu$m and
the $K_{s}$-band, respectively. 
The contribution of the baryonic disc to the rotation curve in NGC 4605, 5949, 5963, 6503
and 6689 was computed assuming that the discs are infinitesimally thin and that 
the stellar mass-to-light ratios, $\Upsilon_{\star}$, in each band are constant with radius. 
For NGC 3621, we followed dB08 who took into account radial $\Upsilon_{\star}$ variations
within the stellar disc and 
assumed a vertical scale height of the stellar disc of one fifth the radial scale length.

In all the galaxies of our sample, the contribution of the gas disc to the rotation
curve is small; at those radii under consideration, its effect is similar 
to a $20\%$ change in $\Upsilon_{\star}$ \citep{bol02,sim03} and, moreover,
this decrease could be partly offset by allowing the stellar disc to have a finite thickness 
(S05). In particular, the contribution of the gas to the rotation curve is $\la 20$
km s$^{-1}$ at the last measured radius ($4$ kpc) in NGC 5963 \citep{bos88}.
Only beyond $10$ kpc, the contribution of the gas is non-negligible as compared to the contribution
of the stellar disc. \citet{ada14} estimated that the contribution of the gas to the NGC 5949
rotation curve is $< 20$ km s$^{-1}$ if the gas distribution follows the stellar distribution.

\begin{figure*}
 \centering\includegraphics[width=15.0cm,height=15.0cm]{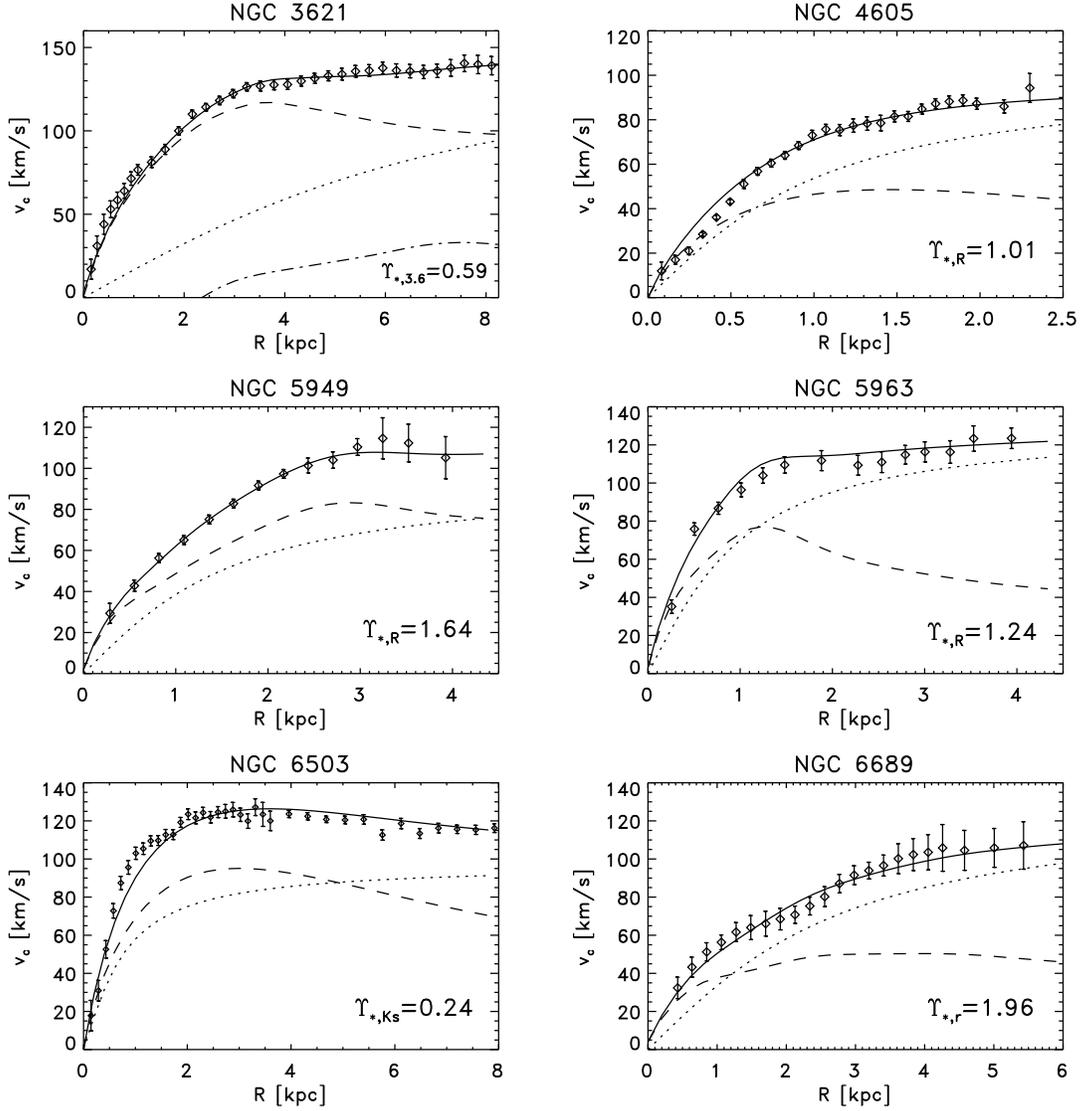}
 \caption{Fits to the observed rotation curves  
using a pseudo-isothermal profile for the dark matter halo. The dashed 
line represents the contribution to the rotation curve of the stellar disc 
using $\Upsilon_{\star}^{\rm syn}$ given in Table \ref{table:DMvalues}. The dotted line shows
the contribution of the dark halo and the dot-dashed line indicates the gas contribution.
}
 \label{fig:RCs_iso}
 \end{figure*}

Figure \ref{fig:RCs_iso} shows the fits to the rotation curves, assuming that the dark matter
haloes follow a pseudo-isothermal sphere, whose profile is
\begin{equation}
\rho_{\scriptscriptstyle \rm ISO}(r)=\frac{\rho_{0}}{1+(r/R_{c})^{2}},
\end{equation}
where $\rho_{0}$ is the central density of the halo and $R_{c}$ the core radius.
Although cosmological simulations predict NFW profiles, it is likely that fluctuations in
the gravitational potential caused by bursts of star formation
in low-mass galaxies are able to transform initial NFW profiles into cored profiles 
\citep[e.g.,][]{pon14}. 
For NGC 3621, 4605, 5949, 5963 and 6689, we adopt the values of $\Upsilon_{\star}$
derived by dB08 and S05 using the colour-$\Upsilon_{\star}$ relations as predicted
in population synthesis models and denoted by $\Upsilon_{\star}^{\rm syn}$
(see Table \ref{table:DMvalues}). For NGC 6503, we choose
$\Upsilon_{\star, Ks}=0.24$, for reasons that will become clear later in \S \ref{sec:RCs_MOND}.
Hence the fits have two free parameters ($\rho_{0}$ and $R_{c}$). The best-fitting parameters were 
reported in dB08 and S05. Although the fits are satisfactory in general, a more cuspy profile
than the pseudoisothermal distribution leads to better fits in NGC 5963 and NGC 6503.
According to Figure \ref{fig:RCs_iso},
NGC 4605, 5963 and 6689 are dark matter dominated galaxies.
In NGC 3621, 5949 and 6503, the contribution of the dark matter halo to the rotation curve
is comparable to the contribution of the baryonic disc, at least at the galactocentric distances
under consideration. In these mass models, NGC 3621 and NGC 4605 have maximum disc
mass-to-light ratios. We need to see if the adopted $\Upsilon_{\star}$-values are consistent
with bar-stability constraints. This will be done in the next Section.

\begin{table}
\begin{minipage}{83mm}
\caption[]{Summary of stability parameters using a pseudo-isothermal dark matter halo.
The suggested range for $\Upsilon_{\star}$ in column 2 comes from dB08 and S05.
In the case of NGC 3621, the two extreme values in the range corresponds to Kroupa and
diet-Salpeter initial mass functions.
The asterisk indicates that the value used was not derived from the colour-$\Upsilon_{\star}$ relation.
}
\vspace{0.01cm}

\begin{tabular}{c c c c c c}\hline
{Name} &  Suggested & Adopted  & {$\left<X_{2}\right>$} & $\Upsilon_{\star}^{\rm (ENL)}$ & 
$\Upsilon_{\star}^{\rm (PWC)}$      \\
{NGC} & $\Upsilon_{\star}^{\rm syn}$ range &$\Upsilon_{\star}^{\rm syn}$ & using  $\Upsilon_{\star}^{\rm syn}$&   &    \\
\hline
3621 &  $0.42-0.59$ & $0.59$& $1.2$ & $0.27$ & $0.25$\\
4605 & $0.94-1.09$ & $1.01$ & $2.9$ & 1.14   & 1.06\\
5949 &  $1.48-1.80$ & $1.64$ & $2.0$ & $0.96$ & 1.24 \\
5963 &  $1.09-1.38$ & $1.24$ & $2.3$ & 1.00 & 1.00 \\
6503 &  --& $0.24^{\ast}$ & $1.9$ & 0.13 & 0.16  \\
6689 & $1.96$ & $1.96$ & $3.3$ & 2.92  & 2.50  \\

\hline
\end{tabular}

\label{table:DMvalues}
\end{minipage}
\end{table}

\begin{figure}
 \centering\includegraphics[width=8.7cm,height=6.5cm]{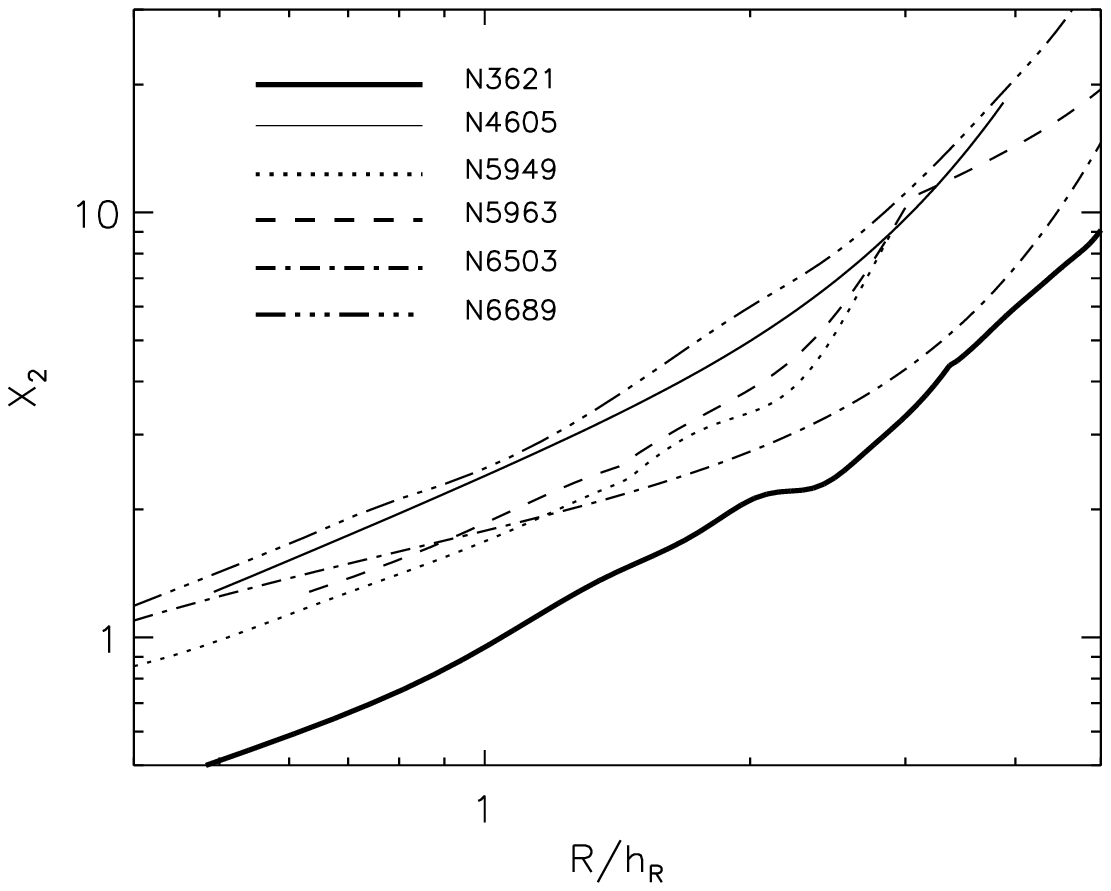}
 \caption{Radial profile of $X_{2}$ for those Newtonian mass models (stellar 
disc plus pseudo-isothermal dark halo) shown in Figure \ref{fig:RCs_iso}.
}
 \label{fig:X2_isothermal}
 \end{figure}

\subsection{Bar-stability constraints on $\Upsilon_{\star}$}
We computed $\Upsilon_{\star}^{\rm (ENL)}$ by fitting
the rotation curve of each galaxy with progressively lower
values of $\Upsilon_{\star}$ until the condition ${\mathcal{R}}=1.1$
was met. The value of $\Upsilon_{\star}^{\rm (ENL)}$ for the
galaxies in question are listed in Table \ref{table:DMvalues}.
We see that $\Upsilon_{\star}^{\rm syn}>\Upsilon_{\star}^{\rm (ENL)}$
in four galaxies and, according to the ENL criterion, they would be
bar unstable when assuming $\Upsilon_{\star}^{\rm syn}$, which
is the mean value derived from population synthesis models. However, only
for two galaxies (NGC 3621 and NGC 5949), $\Upsilon_{\star}^{\rm (ENL)}$
is well below the range suggested by population synthesis models.

In order to compute $\Upsilon_{\star}^{\rm (PWC)}$, we have derived
$X_{2}$ versus $R$ for our six galaxies under consideration, and
for the mass models presented in Section \ref{sec:RCs_DM} (having $\Upsilon_{\star}$-values
as indicated by galaxy colours). The radial profile of $X_{2}$ is shown
in Figure \ref{fig:X2_isothermal}, where the radius is normalized in terms of the scale length
of the disc $h_{R}$. Since the radial brightness profile of the disc in some galaxies 
is not perfectly exponential, we have defined $2h_{R}$ as the galactocentric distance where
the mass surface density drops a factor of $\exp(2)$ from 
its central value. We see that NGC 3621 exhibits the lower values 
of $X_{2}$.

Since the four S05 galaxies and NGC 3621 exhibit no indication of a strong bar, we may
conservatively assume that for these five galaxies it holds that $\left<X_{2}\right> > 2.7$.
The resultant $\Upsilon_{\star}^{\rm (PWC)}$ values in our galaxies are given in 
Table \ref{table:DMvalues}.

We see that $\Upsilon_{\star}^{\rm (ENL)}$ and 
$\Upsilon_{\star}^{\rm (PWC)}$ are not dissimilar; it holds that 
\begin{equation}
0.85\leq \frac{\Upsilon_{\star}^{\rm (PWC)}}{\Upsilon_{\star}^{\rm (ENL)}}\leq 1.3.
\end{equation}
For the five galaxies with empirical determinations of $\Upsilon_{\star}^{\rm syn}$ 
from galaxy colours,
$\Upsilon_{\star}^{\rm (ENL)}$ and $\Upsilon_{\star}^{\rm (PWC)}$
are significanly smaller than those values predicted from population synthesis models in two
galaxies: NGC 3621 and NGC 5949. 
The $\Upsilon_{\star}^{\rm syn}$-value in NGC 3621 is a factor
of $1.5$-$2$ larger than the upper value inferred to have bar stability. However, several assumptions
have to be made to compute $\Upsilon_{\rm syn}$, such as the choice of the initial stellar mass
function or the star formation history, which may account for this discrepancy \citep[see][for a review]{con13}.
For NGC 5949, values of $\Upsilon_{\star,R}$ larger than $\sim 1.3$ 
are not permitted from the stability analysis. Interestingly, \citet{ada14} derived the best-fitting
value of $\Upsilon_{\star,r}$ in NGC 5949 from kinematic models and found that it is $1.16\pm 0.34$ 
when gas-based models are used, and $1.20\pm 0.28$ when the stellar kinematics is 
used instead.

The stability of one galaxy of our sample, NGC 6503, has been examined in
great detail in the literature. 
\citet{bot97} concluded that the peak of the contribution of
the disc to the rotation curve cannot contribute
more than $90$ km s$^{-1}$ in NGC 6503 because the disc would develop 
a prominent (unobserved) bar. This condition implies that $\Upsilon_{\star,Ks}\leq 0.21$.
The knowledge of the line-of-sight velocity dispersion
profile allows a more sophisticated analysis. \citet{bot97}
found that the relatively low values of the observed velocity dispersion
are reproduced in the simulated galaxy when the disc contributes $\sim 70$ km s$^{-1}$
at its peak (or, equivalently, $\Upsilon_{\star,Ks}\leq 0.13$). Therefore,
the values of $\Upsilon_{\star}^{\rm(ENL)}$ and $\Upsilon_{\star}^{\rm(PWC)}$ obtained 
for NGC 6503 are consistent with these previous estimates.

In summary, we find that (1) $\Upsilon_{\star}^{\rm (ENL)}$ and $\Upsilon_{\star}^{\rm (PWC)}$
are not dissimilar and (2) in two of the galaxies 
the values for $\Upsilon_{\star}$ suggested by population-synthesis models are a factor of
$1.5-2$ larger than those values allowed by bar-stability arguments.

\subsection{A note on the spiral arm multiplicity}
\label{sec:note}
The number of spiral arms may also set constraints on $\Upsilon_{\star}$.
This kind of analysis is adequate for galaxies having low inclinations 
in order to trace out correctly the spiral arms \citep[e.g.,][]{don15}.
As discussed in \S \ref{sec:sample}, some of the galaxies in our sample exhibit a 
multi-armed shape (NGC 3621 and NGC 5963).
In particular, NGC 5963, the galaxy with the less inclination in our sample, 
shows the presence of four tightly wound spiral arms.
Thus, the surface density in this galaxy should be small enough for the
the amplification of bisymmetric ($m=2$) modes to be highly suppressed.
Assuming $\alpha=1.5$ (see \S \ref{sec:spiralarms}) and a stellar disc with 
$\Upsilon_{\star}=1.0$ (see previous Section), the $m=4$ mode is efficiently 
swing amplified from $0.8 h_{R}$ (or $\sim 0.55$ kpc) to $1.5h_{R}$ (or $1.0$ kpc).
This dominance of the $m=4$ mode may be consistent with the spiral morphology
seen in the optical images (see Figure \ref{fig:images2} in the
present paper and Figure 1c in S05), although the $3.6\mu$m image shows that
two arms are more prominent than the other two.

\begin{figure*}
 \centering\includegraphics[width=15.0cm,height=15.0cm]{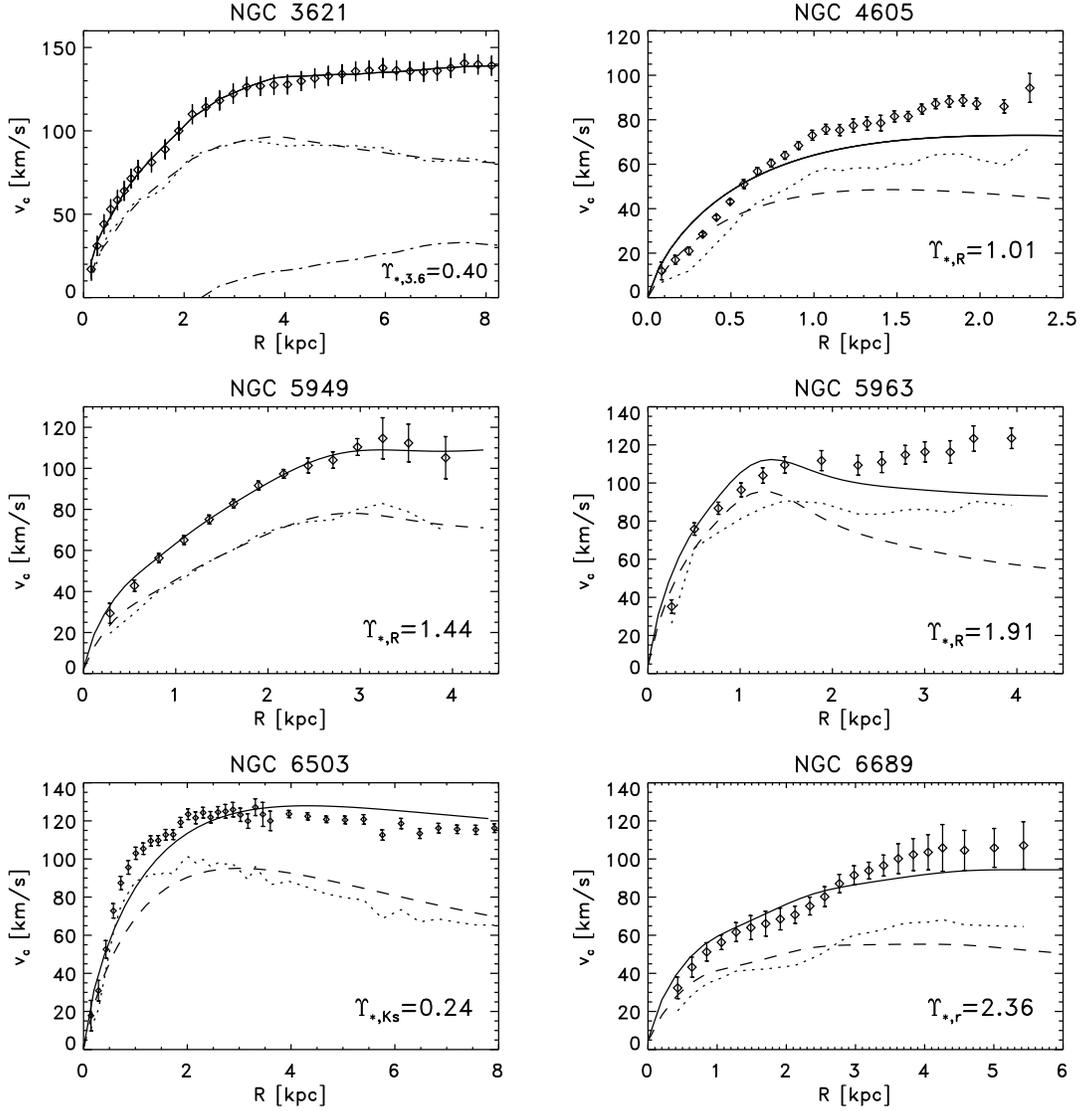}
 \caption{MOND fits to the observed rotation curves using the simple $\mu$-function
with $a_{0}=1.2\times 10^{-8}$ cm s$^{-2}$ and treating 
$\Upsilon_{\star}$ as free parameter, which is quoted
at the right corner at each panel. The dashed curves represent the Newtonian 
contribution of the stellar disc. The dotted lines indicate the contribution of the
stellar disc required by MOND in order to have a perfect fit to the observed rotation 
curve.  
}
 \label{fig:RCs_a012}
 \end{figure*}

\section{Analysis in MOND}
\label{sec:MOND_analysis}
\subsection{MOND fits to the rotation curves}
\label{sec:RCs_MOND}

Fits to the rotation curves of NGC 3621 using MOND can be found in \citet{gen11},
\citet{ang12} and \citet{ran14}.
%Gentile et al. (2011), Angus et al. (2012) and Randriamampandry \& Carignan (2014). 
These authors used the same THINGS data
as we adopt here. The rotation curve of NGC 6503 was studied in the MOND context by
\citet{bot15}.

In both formalisms described in Section \ref{sec:MONDformalisms} (the non-linear Poisson equation 
and QUMOND), the following algebraic relation between the `true' acceleration $g_{M}$ and 
the Newtonian acceleration $g_N$ (created by the visible components)
is a good approximation in the midplane of smooth axisymmetric disc systems  
%(Milgrom 1986; Angus et al. 2012)
\begin{equation}
\mu\left(\frac{|\vecg_{M}|}{a_{0}}\right)\vecg_{M}=\vecg_{N}
\label{eq:algMOND}
\end{equation}
\citep{mil86,ang12}.
Equation (\ref{eq:algMOND}) relates
the MOND circular velocity $v_{c,M}=(g_{M}R)^{1/2}$ to the Newtonian circular
velocity $v_{c,N}\equiv (g_{N}R)^{1/2}$ as
\begin{equation}
\mu(x)v_{c,M}^{2}=v_{c,N}^{2},
\label{eq:algebraic}
\end{equation}
with 
\begin{equation}
x=\frac{v_{c,M}^{2}}{Ra_{0}}.
\label{eq:def_x}
\end{equation}

\begin{table}
\begin{minipage}{83mm}
\caption[]{Summary of stability parameters using MOND. The quoted intervals in $\Upsilon_{\star}^{\rm fit}$
and $\left<X_{2}\right>$ correspond to the $95\%$ posterior credible bands.
The range for $\Upsilon_{\star}^{\rm syn}$ in column 2 comes from dB08 and S05.
In the case of NGC 3621, the two extreme values in the range corresponds to Kroupa and
diet-Salpeter initial mass functions.
}
\vspace{0.01cm}

\begin{tabular}{c c c c c}\hline
{Name} &  $\Upsilon_{\star}^{\rm syn}$ range & {$\Upsilon_{\star}^{\rm fit}$} & ${\mathcal{R}}$ & 
$\left<X_{2}\right>$      \\
{}&   & & & using  $\Upsilon_{\star}^{\rm fit}$   \\
\hline
NGC 3621 & $0.42-0.59$ &   $0.401\pm 0.014$    & $0.92$ & $1.39\pm 0.04$\\
NGC 4605 &   $0.94-1.09$ & $1.010\pm^{0.066}_{0.056}$ & $0.96$  & $1.57\pm^{0.08}_{0.09}$\\
NGC 5949 &   $1.48-1.80$ & $1.444\pm^{0.146}_{0.143}$ & $0.87$ & $1.70\pm^{0.17}_{0.14}$ \\
NGC 5963 &   $1.09-1.38$ & $1.910\pm^{0.241}_{0.255}$ & $0.73$ & $1.32\pm^{0.19}_{0.14}$\\
NGC 6503 &     --         & $0.24\pm{0.012}$ & $0.82$ & $1.47\pm^{0.07}_{0.06}$  \\
NGC 6689 &   $1.96$ & $2.366\pm^{0.455}_{0.424}$ &  $1.10$ & $1.84\pm^{0.36}_{0.27}$ \\

\hline
\end{tabular}
%\medskip\\

\label{table:MONDvalues}
\end{minipage}
\end{table}

\citet{fam05} found that the ``simple'' $\mu$-function, $\mu=x/(1+x)$,
fits the data of NGC 3198 and the terminal-velocity curve of the Milky Way much more
satisfactorily than the standard interpolating function. 
\citet{fam07} and \citet{san07} also found that the
simple $\mu$-function provides more plausible values 
of the relative stellar mass-to-light ratios for bulge and disc, as well as
the generally smaller global mass-to-light ratio, for a sample of galaxies having a gradual transition
from the Newtonian limit in the inner regions to the MOND limit in the outer parts.
For the galaxies in the sample of \citet{gen11},
the simple interpolating function also yields better fits to the rotation curves than the 
standard interpolating function \citep[see also][]{wei08}.
%(see also Weijmans et al.~2008).

\begin{figure}
 \centering\includegraphics[width=8.75cm,height=6.5cm]{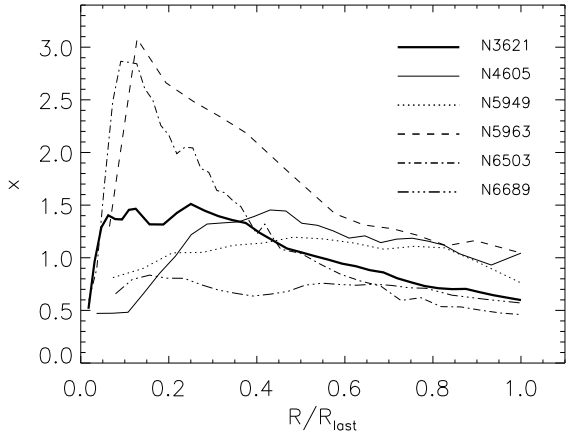}
 \caption{Observed radial acceleration divided by $a_{0}$ (see Equation \ref{eq:def_x}) versus $R$ for each galaxy.
$R_{\rm last}$ is the radius of the last radial point in Figure \ref{fig:RCs_a012}.
}
 \label{fig:x_a012}
 \end{figure}

Figure \ref{fig:RCs_a012} shows the fits to the rotation curves using
the MOND framework with the simple interpolating $\mu$-function. 
The Newtonian circular velocities $v_{c,N}$ (without dark matter) were determined
as described in \S \ref{sec:RCs_DM}. We made fits
with $a_{0}=1.2\times 10^{-8}$ cm s$^{-2}$ and only $\Upsilon_{\star}$ as a
free parameter. The required stellar mass-to-light ratios in the four S05 
galaxies obtained in the fit to the rotation curves ($\Upsilon_{\star}^{\rm fit}$) are given
in Table \ref{table:MONDvalues}. The intervals shown in this Table correspond to
$95\%$ posterior credibility intervals obtained using the Bayesian inference method
presented in Appendix \ref{sec:statistics}. The $\Upsilon_{\star}^{\rm fit}$ values 
are consistent with those derived using the colour-$\Upsilon_{\star}$ relations 
predicted in population synthesis models except for NGC 5963 (see Table \ref{table:MONDvalues}). 
The relatively large values of $g_{M}/a_{0}$ in NGC 5963 yield a 
$\Upsilon_{\star}^{\rm fit}$-value close to the maximum disc mass-to-light ratio.

We see that the fit is good for NGC 3621 and NGC 5949, acceptable for NGC 6689,
but poor for NGC 4605, 5963 and 6503. 
In order to highlight the significance of the MOND fits, the dotted lines in Figure
\ref{fig:RCs_a012} show the radial profiles of the Newtonian circular velocity required to match
the observed rotation curve with MOND. 
It is a complementary way to quantify what MOND requires to provide
a successful fit. We see that for NGC 4605 and NGC 5963,
the discrepancy between the required and the observed Newtonian
circular velocities is significant.

\begin{figure}
 \centering\includegraphics[width=8.5cm,height=6.0cm]{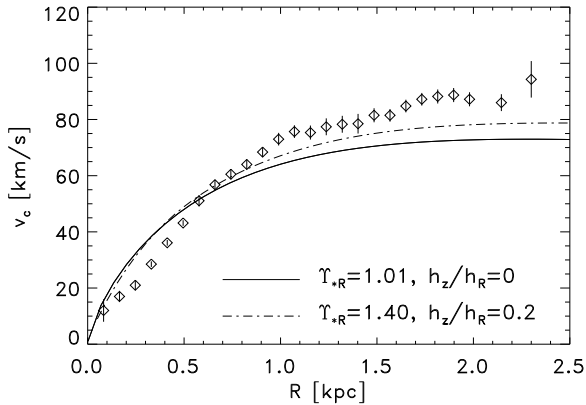}
 \caption{The difference in the MOND rotation curves for NGC 4605 in the case
of null thickness (solid line) and a finite thickness (dot-dashed line)
of the stellar disc.
}
 \label{fig:thickness_NGC4605}
 \end{figure}

Although $a_{0}$ and the interpolating function should be universal, they must
be determined empirically. Interestingly, in galaxies having $x\ga 1$ at every observed 
radius, or galaxies with $x$ constant along $R$, the shape of the
rotation curve predicted by MOND is barely sensitive to the particular choice of the 
interpolating function or on the adopted value of $a_{0}$. According to Equation (\ref{eq:algebraic}),
for those galaxies, MOND predicts that the rotation curve should be a scaled version
of the Newtonian rotation curve.  
Figure \ref{fig:x_a012} shows $x= v_{c}^{2}/(Ra_{0})$,
with $v_{c}$ the observed rotation curve. We see that for NGC 5949 and NGC 6689, $x$
is approximately constant with radius. For NGC 5963, $x\ga 1$ at all the observed
galactocentric distances. Thus, for these three galaxies, the fit to the rotation
curves are expected to be insensitive to small changes to the adopted value of
$a_{0}$ or the exact form of the interpolating function.
In fact, we recalculated the fits for $a_{0}=0.9\times 10^{-8}$ cm s$^{-2}$ and for
$1.2 \times 10^{-8}$ cm s$^{-2}$, leaving $\Upsilon_{\star}$ as the free parameter.
Using these values for $a_{0}$, the differences in the circular velocity are less than $3\%$,
making them barely distinguishable.

We have explored other values of the index $n$ in the interpolating function (see Eq. \ref{eq:mufamily}), 
and found that $\Upsilon_{\star}$ does
depend on the value of $n$, but the quality of the fits to the rotation curves are rather insensitive.
None $n$-value provides satisfactory fits to the rotation curves of all the galaxies
in our sample.  Since $n=1$ provides $\Upsilon_{\star}$-values very close to 
what the colour-$\Upsilon_{\star}$ relation predicts \citep[see also][]{fam07},
we restrict our analysis to the simple $\mu$-function in the remainder of the paper.

Distances to the galaxies are another source of uncertainty (see Table \ref{table:basic_parameters} 
for the estimated uncertainties in the distances). 
It is easy to show that it is equivalent to vary $D$ by a factor of $\alpha$ (with $a_{0}$ fixed) than changing
$a_{0}$ by the same factor $\alpha$ (with $D$ fixed). We find that the goodness of the fits to the rotation
curves is almost unaltered if the distance to these galaxies is allowed to vary by $15$ percent.

The uncertainty in the distance of NGC 5949 could be significantly
larger than $15\%$. For this galaxy, S05 derived
a distance of $14.0\pm 2.4$ Mpc from the Tully-Fisher relation. 
\citet{ada14} reported a distance of $14.3$ Mpc implied by the Hubble flow, 
whereas \citet{spa08} assumed a distance of $7.4$ Mpc. Fortunately,
for this galaxy, $x$ is approximately constant with galactocentric radius
and hence the goodness of the fit to the rotation curves is not very sensitive
to the adopted distance.

Figures \ref{fig:RCs_a012}  indicates that in the central regions 
of NGC 4605, 5963 and 6689,
MOND predicts higher rotation speeds than observed. A slightly better fit to the 
rotation curves of these galaxies can be achieved by including the finite thickness 
of stellar discs. Figure \ref{fig:thickness_NGC4605} shows
the MOND rotation curve for NGC 4605 with $\Upsilon_{\star,R}=1.4$, when
the Newtonian acceleration at the midplane was calculated assuming that the stellar
scale height is one fifth the stellar scale length. The fit is still not fully
satisfactory; the MOND fit within $R=2.4$ kpc predicts an asymptotic value of 
$80$ km s$^{-1}$, which is significantly below the circular velocity of $100$ km s$^{-1}$
measured by \citet{rub80} at $3.3$ kpc.  However, uncertainties in the ellipticity
caused by the non-axisymmetric structures seen in both photometric and kinematic maps 
(see Figure \ref{fig:images1} and S05) might be the reason why MOND does not correctly predict the rotation curve 
of NGC 4605.

MOND is also unable to reproduce the shape of the rotation curve of NGC 5963; 
the observed rotation curve at the peak of the disc is still rising, but MOND
predicts that it should slowly drop in a similar fashion as the Newtonian circular velocity does,
because $x>1$. This feature is also present in other galaxies.
For instance, the compact galaxies UGC 5721, 7399, 8190 and 9179 also exhibit a rising of 
the rotation curve beyond the turnover of the Newtonian rotation curve 
\citep{spa08,swa10},
%(Spano et al. 2008; Swaters et al. 2010), 
which MOND cannot reproduce \citep[e.g.,][]{swa10}.
Although the rotation curves of UGC 5721, 7399 and 8490 
were classified as high quality, based on asymmetry, bar, and so on,
Swaters et al (2010) identified twisting isophotes in the central regions of UGC 5721
(suggesting the presence of non-circular motions),
the bar in UGC 7399, and the large warp in UGC 8490 as possible agents which may affect
the derived rotation curve. 
The galaxy NGC 5963 also contains non-circular motions of the
order of $\sim 15$ km s$^{-1}$ (S05), which may affect the observed rotation curve 
at $R>3$ kpc. 

Moreover, the peculiar surface brightness map of NGC 5963, showing a steep decline 
beyond $\sim 1$ kpc, may indicate that the assumption of a constant $\Upsilon_{\star}$
across this galaxy is not a good approximation. The surface brightness profile may
be modelled by the sum of two exponential components, each of them having different 
$\Upsilon_{\star}$. Using a double exponential profile, the MOND fit to the rotation
curve is slightly improved.

In NGC 5963, the asymptotic velocity predicted using our MOND fit to the rotation 
curve is  $98$ km s$^{-1}$, small as compared to the asymptotic velocity 
measured by \citet{bos88}. However, we have neglected the gas mass,
which is significant at larger radii. By including the gas mass, the asymptotic circular
velocity expected under MOND, $(GMa_{0})^{1/4}$, is larger.

In NGC 6503, MOND underpredicts the circular speed at the radial interval $0.8$ kpc $<R< 2.5$ kpc. 
If the dust obscures part of the light, this discrepancy may be expected because of
our poor correction for dust extinction. \citet{bot89}
suggested that dust extinction could be responsible for the plateau observed
between $R=0.1$ kpc and $R=1$ kpc in $B$- and $R$-band surface brightness profiles.
For NGC 6503, we have made a fit with both $\Upsilon_{\star,Ks}$
and the scale length of the exponential stellar disc $h_{R}$, as free parameters. The best fitting
parameters were found to be $\Upsilon_{\star,Ks}=0.35\pm 0.017$
and $h_{R}=1.06\pm 0.03$ kpc. Nevertheless,
this hypothesis that the light distribution does not trace the underlying
mass distribution due to the presence of dust, has been ruled out
by \citet{kuz12} on the basis that the plateau is also
present in the $K_{s}$-band photometry and on the agreement between H$\alpha$
and CO rotation curves, which suggests that extinction is minimal.

\subsection{The $X_{m}$-parameter in the MOND framework}
\label{sec:XmMOND}

\begin{figure*}
 \centering\includegraphics[width=15cm,height=15cm]{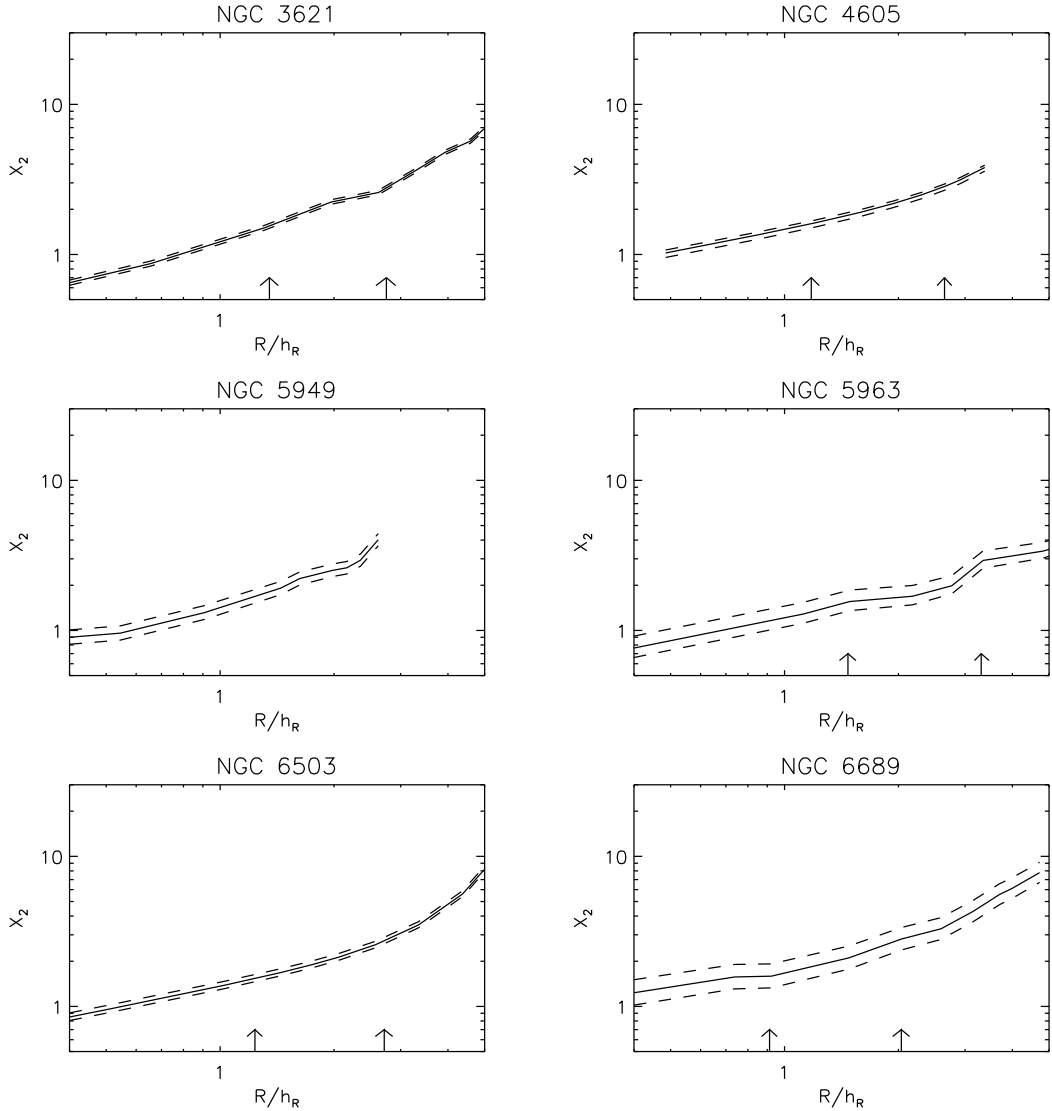}
 \caption{$X_{2}$-parameter versus $R$ under MOND (solid 
line) together with the $95\%$ posterior credible band (dashed lines). We have used the
simple $\mu$-function with $a_{0}=1.2\times 10^{-8}$ cm s$^{-2}$.
The arrows indicate the maximum galactocentric distances where the swing amplification of
the modes $m=2$ (left arrow) and $m=4$ (right arrow) can occur in each galaxy (except for NGC 5949).
}
 \label{fig:X2_MOND}
 \end{figure*}

Figure \ref{fig:X2_MOND} shows the MOND $X_{2}$ parameter as a function of $R$ for the galaxies 
in our sample, when the simple $\mu$-function is used: $\mu(x)=x/(1+x)$ and $L(x)=1/(1+x)$.
We see that the values of $X_{2}$ at a given $R/h_{R}$ vary little from galaxy to galaxy.

In Section \ref{sec:RCs_MOND}, we found that the MOND fits to the rotation curves
for NGC 4605 and NGC 5963 are clearly unsatisfactory and thus any further
analysis will be relevant once we can identify the cause
of this discrepancy. If the explanation for the poor fit to the rotation curve is
only that the measured rotation curves for these galaxies are affected by the presence 
of non-circular motions in the gas, we may put aside the rotation curves from the analysis, and recalculate
$X_{2}$ for reasonable values of $\Upsilon_{\star}$. Doing so, we find that the radial
profile of $X_{2}$ is rather insensitive to the adopted value of $\Upsilon_{\star}$ (see
also Section \ref{sec:bar_ins_MOND}). For instance, even if we adopt a value twice lower 
than $\Upsilon_{\star}^{\rm syn}$, the $X_{2}$ parameter increases by less than $15\%$; the effect is 
not significant.  The reason is that adopting a smaller $\Upsilon_{\star}$ leads to a smaller surface
density but also to a slower rotation and thereby a lower $\kappa$, which all together
results in a similar $X_{2}$.

As a second possibility, we may assume that the measured rotation curves and the quoted
error bars are reliable but the mass in the disc of NGC 4605 and NGC 5963 does not follow 
light (or the photometry analysis is not precise enough to derive the distribution
of stellar mass). If MOND is correct, 
we can derive both the disc surface density and $X_{2}(R)$ from the observed 
rotation curve alone. 
Doing this exercise for NGC 4605, we found that $X_{2}$ increases
by $11\%$ at a galactocentric distance of $2h_{R}$ (or $1.36$ kpc), and by $30\%$ 
at $3h_{R}$ (i.e. $2$ kpc).
Since the PWC criterion for bar stability depends on the value of $X_{2}$ within $2h_{R}$,
the $11\%$ enhancement in $X_{2}$ is again not significant as far as bar-stability
analysis concerns.

As a summary of this Section, we conclude that for the galaxies in our sample, 
the profiles of $X_{2}$ vs $R/h_{R}$ are rather similar from galaxy to galaxy.
The low values of $X_{2}$ in MOND are intrinsic to the basic tenets of MOND.
For the galaxies with poor MOND fits to the rotation curves, we have explored other
scenarios but find that the MOND values of $X_{2}$ are rather robust to observational
uncertainties.
The analysis for NGC 5963 should be taken with caution because this galaxy
has an unusual surface brightness, which may indicate that a constant $\Upsilon_{\star}$
may not be a good approximation.

\subsection{A short comment on arm multiplicity in MOND}
In Figure \ref{fig:X2_MOND}, we have also marked the predicted extension of the 
spiral arms with multiplicity $m=2$ and $m=4$ by using that $m=2\alpha^{-1}X_{2}$
and $\alpha\simeq 1.25$ for all the galaxies except for NGC 5949, because it 
has an almost linear rotation curve (see \S \ref{sec:spiralarms}).
In the case of NGC 5963, the galaxy discussed 
in \S \ref{sec:note} because its low inclination allows us to trace out their spiral arms, 
the swing amplification theory predicts that, under MOND, it should 
show two spiral arms within $1.3 h_{R}\simeq 0.9$ kpc. 
From the present analysis, it is not easy to decide if the observed spiral arm multiplicity
is consistent with this prediction or not; S05 argued that this galaxy contains four spiral arms in
detailed optical images (see figure 1c in S05), which is inconsistent with the MOND prediction 
that it should exhibit only two arms.
On the other hand, the $3.6\mu$m image in Figure \ref{fig:images2} shows two prominent spiral arms
but also two fainter spiral arms. 
We remind that the spiral wave theory is implicitely assuming that discs are relatively cold 
(see \S \ref{sec:spiralarms}). In hot discs, the self-gravity of the disc is
weaker and the amplification factor of spiral waves becomes lower. This possibility
will be discussed in \S \ref{sec:hotdisks}.

\subsection{Bar instability in MOND}
\label{sec:bar_ins_MOND}

We first consider the ENL stability criterion.
The values of ${\mathcal{R}}$ for the six galaxies in MOND are listed in Table \ref{table:MONDvalues}.
For all galaxies except NGC 6689, ${\mathcal{R}}$ is less than the
critical value $1.1$. This implies that these five galaxies are bar unstable not only in
the MOND context, but also in the equivalent Newtonian systems. In other words,
the values of $\Upsilon_{\star}$ required to account for the rotation curves
in MOND are so large and the discs so massive that these galaxies would be unstable 
against bar modes even under Newtonian dynamics.

\begin{figure*}
 \centering\includegraphics[width=15cm,height=15.0cm]{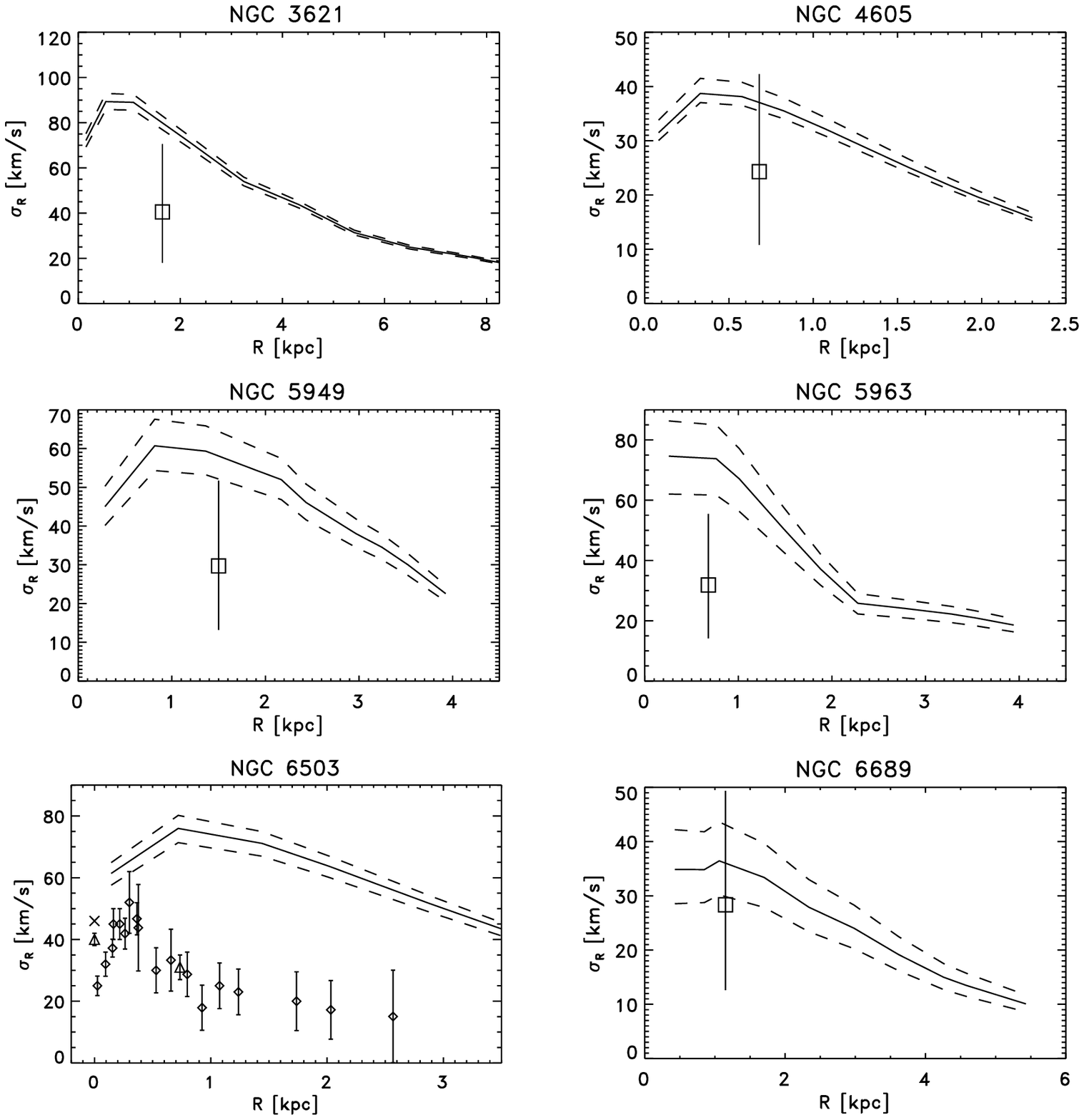}
 \caption{Radial velocity dispersions in order to have a constant Toomre parameter $Q_{M}=2.2$ in MOND.
The dashed lines show the $95\%$ posterior credible interval.
In the case of NGC 6503, the observed line-of-sight velocity dispersion is also
shown from \citet{bot89} (diamonds with error bars), \citet{bar02} (cross at $R=0$)
and \citet{kor10} (triangles with error bars). For the remainder four galaxies,  we plot the 
expected $\sigma_{R}$ at one scalelength (open square), as derived using a sample of 
$62$ late-type spiral galaxies with published stellar velocity dispersions (see \S \ref{sec:hotdisks}).
The vertical bar indicates the range that contains $95\%$ of the galaxies in
these sample of $62$ galaxies.
}
 \label{fig:velocity_dispersion_stability}
 \end{figure*}

  We have computed $\left<X_{2}\right>$ in MOND. In Table \ref{table:MONDvalues},
we provide the values of  $\left<X_{2}\right>$ and also the $95\%$ posterior credibility
interval obtained using the Bayesian approach. The value of $\left<X_{2}\right>$ is very 
insensitive to changes in the adopted inclination angle of the galaxy, distance or observational
uncertainties in the rotational velocity. 
For the galaxies in our sample, $\left<X_{2}\right>$ lies between $1.4$ and $1.9$ 
(see Table \ref{table:MONDvalues}). Accordingly, the PWC criterion predicts that all 
the galaxies should have strong bars, unless the stellar discs are dynamically hot. 

Due to the relatively high inclination of NGC 4605, we cannot rule out
the existence of bars in this galaxy. One could argue that the noncircular motions
detected in NGC 4605 are due to a hidden bar. If the bar is aligned with the major axis, 
the rotational velocities using standard tilted-ring models are underestimated \citep[e.g.][]{ran15}
and this could explain why the MOND fit to the rotation curve in NGC 4605 is so poor. A more 
detailed analysis using different wavelengths, as that done for NGC 2976 in \citet{val14}, 
is required to test if this interpretation is correct.

Since there is no evidence for strong bars in any of the galaxies in our sample (see \S \ref{sec:sample}), 
it is worthwhile to explore a scenario where galaxies are stable due to their stellar random motions. 
This possibility is not new and has been 
discussed in \citet{sel16} in a general context. He found that this scenario is unsatisfactory in the case 
of late-type spirals with stellar masses between $10^{8}M_{\odot}$ and several $10^{9}M_{\odot}$, 
because these galaxies do not show evidence of strong dynamical heating as far as they do not 
possess bulge, they are rather thin and display spiral-arm structure, which indicates that they are not
so hot to suppress the growth rates for bar instabilities.

\subsection{Supression of bar instabilities: dynamically-hot discs?}
\label{sec:hotdisks}
In this Section, we discuss the magnitude of the velocity dispersion
to provide stability. 
\citet{ath86} found that isolated (two-dimensional) Kuzmin-Toomre discs 
with $\left<Q\right>\geq 2.2$ are stable to bar instability. In the dynamical models of 
NGC 6503, \citet{pug10} also found that a weak bar is developed in models where 
the minimum $Q$-value is $\simeq 2.2$. 

For each galaxy, we have computed the radial velocity dispersion as 
a function of $R$ assuming that the MOND Toomre parameter is constant
and equal to $2.2$ at any galactocentric distance. We have assumed
that the galaxies are purely stellar discs. The radial profiles
of $\sigma_{R}$ are shown in Figure \ref{fig:velocity_dispersion_stability}. 
They should be interpreted as the minimum radial velocity dispersions required 
to guarantee disc stability under MOND theory.

For the galaxies in the present sample, there are measurements of the
light-of-sight velocity dispersion only for NGC 6503 \citep{bot89,bar02,kor10}.
For galaxies with high inclinations as 
NGC 6503, the line-of-sight velocity dispersion is given by a weighted combination of the 
radial ($\sigma_{R}$) and tangential ($\sigma_{\phi}$) components 
of the velocity dispersion. 
Using that $\sigma_{\phi}\simeq 0.7\sigma_{R}$ from the epicyclic approximation and
following \citet{ber10}, it is easy to infer that  the line-of-sight
velocity dispersion, in highly inclined galaxies, is smaller than the radial velocity dispersion 
by $20\%$ at most \citep[see also][]{kre05a}. 
\citet{bot89} observed at $5020\AA$, and the line-of-sight velocity dispersion profile
is shown in the corresponding panel in Figure \ref{fig:velocity_dispersion_stability}. 
\citet{bar02} and \citet{kor10} observed at $\sim 8500\AA$ and obtained a larger value
of the velocity dispersion in the centre of the galaxy (see Figure \ref{fig:velocity_dispersion_stability}). 
The reasons for this discrepancy are still not clear \citep[see][for a discussion]{kor10}, but it is likely
that they are observing different populations at the centre of the galaxy. Anyway, even adopting the 
largest values for the line-of-sight velocity dispersions ($\sim 40$ km s$^{-1}$), NGC 6503 is not 
dynamically hot enough to assess the level of stability required. Why NGC 6503 does not contain a 
strong bar has so far defied explanation within MOND.

For the remainder five galaxies, there is no direct measurements of the velocity
dispersion, but we may use some correlations derived in other galaxies to evaluate
whether the required velocity dispersion to satisfy stability is feasible or not.
In fact, observations indicate that those galaxies with higher circular velocity
present larger stellar velocity dispersion. For a sample of $11$ galaxies, 
\citet{bot93} found a linear trend between the central value of the vertical
velocity dispersion, which we denote by $\sigma_{z}(0)$ (and it is approximately
$\sigma_{R}$ at one scalelength) and the galaxy maximum rotation velocity $v_{\rm max}$
\begin{equation}
\sigma_{z}(0)=\sigma_{R}(h_{R})=(0.29\pm 0.10)v_{\rm max}.
\end{equation}
For a sample of $15$ intermediate to late-type edge-on galaxies,
\citet{kre05b} found a similar correlation 
\begin{equation}
\sigma_{z}(0)=\sigma_{R}(h_{R})=(0.22\pm 0.10)v_{\rm max}+(10\pm 17).
\end{equation}
Moreover, \citet{kre05b} combined  all the galaxies with known stellar disc velocity 
dispersion at that time, gathering a sample of $36$ galaxies, which includes 
from S0-a to Scd galaxies. For that sample of galaxies, they reported that 
\begin{equation}
\sigma_{z}(0)=\sigma_{R}(h_{R})=(0.33\pm 0.05)v_{\rm max}+(-2\pm 10).
\end{equation} 
Finally, for a sample of $30$ close to face-on galaxies, \citet{mar13}
found the following relationship
\begin{equation}
\sigma_{z}(0)=(0.248\pm 0.038)v_{\rm 2.2},
\end{equation} 
where $v_{2.2}$ is the circular velocity at a radius $2.2h_{R}$.

From the above empirical correlations, we infer that galaxies with $v_{\rm max}=100$
km s$^{-1}$, as those galaxies in the present sample, have a mean $\sigma_{R}$ at one 
scalelength of $30$ km s$^{-1}$, with a $1\sigma$ scatter of $10$ km s$^{-1}$.

\begin{figure}
 \centering\includegraphics[width=8.5cm,height=7.0cm]{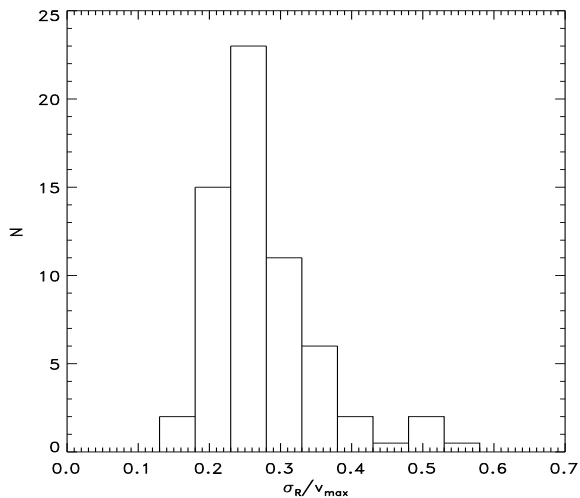}
 \caption{ Histogram of ${\mathcal{B}}$, defined as the ratio between $\sigma_{R}$ at $h_{R}$ 
and $v_{\rm max}$ (the maximum of the observed rotation curve), for a sample of $62$ spiral 
galaxies.
}
 \label{fig:histogram_vel_disp}
 \end{figure}

To have a more complete picture of the full sample of galaxies, we have constructed a 
histogram of the distribution of the parameter ${\mathcal{B}}$, defined as ${\mathcal{B}}\equiv
\sigma_{R}(h_{R})/v_{\rm max}\simeq \sigma_{z}(0)/v_{\rm max}$,
 for the total sample of $62$ galaxies, consisting of the extended
sample of \citet{kre05b} plus the sample in \citet{mar13}, but excluding
the four early galaxies of type S0 and Sab (see Figure \ref{fig:histogram_vel_disp}). 
We find that $95\%$ of the galaxies
have ${\mathcal{B}}$ values in the interval between $0.13$ and $0.45$.
For each galaxy, we have computed this interval for the expected value of $\sigma_{R}$
at $h_{R}$ using the observed value of $v_{\rm max}$ and they are shown in 
Figure \ref{fig:velocity_dispersion_stability}
as an open square and a vertical bar. 

In the case of NGC 4605, we already discussed in \S \ref{sec:bar_ins_MOND} that a more detailed
analysis is required to decide if NGC 4605 has a bar or not. Hence, it remains unclear if this galaxy is 
stable to bar formation.

From Figure \ref{fig:velocity_dispersion_stability}, we see that the expected velocity dispersion 
is fully consistent with the values required to have $Q_{M}=2.2$ in NGC 6689. Remind that for this galaxy, 
there is no any evidence for a bar either photometrically or kinematically. 
Indeed, S05 did not detect any deviations from circular motions. If the
radial velocity dispersion at $1h_{R}$ is confirmed to be larger than $35$ km s$^{-1}$,
MOND could explain satisfactorily the absence of a bar in this galaxy.

However, for NGC 3621, 5949 and 5963 the probability is less than $5\%$
meaning that they require an abnormally large radial velocity dispersion to recover stability.
We must stress that NGC 3621 and NGC 5949 are ideal systems to study their stability and 
can provide a stringent test to MOND because they are representative of pure-disc galaxies with no bars
and MOND reproduces their high-resolution rotation curves fairly well, indicating that the
distribution of mass in these galaxies is well determined in the MOND context.

NGC 5963 is a genuinely puzzling system in MOND; the fit to the rotation curve is poor
and the disc is likely to be unstable to bar modes (unless $\sigma_{R}$ at $1h_{R}$
is larger than $\sim 70$ km s$^{-1}$). It is unclear if a more realistic
modelling, rather than using our 
oversimplifing assumptions (e.g. a constant $\Upsilon_{\star}$ across the galaxy, 
see \S \ref{sec:RCs_MOND}), could bring this galaxy into agreement with observations.

Measurements of stellar velocity dispersions would be of enormous interest.
The galaxies in our sample are suitable for measuring the stellar velocity dispersions
because they have relatively high surface brightness (see Table \ref{table:basic_parameters}).
We need measurements of the velocity dispersions
of old stars because they are the population that more
contributes to the mass budget in the disc and thus the best tracer to
infer the mass-weighted velocity dispersion, which is required to compute the Toomre parameter.
\citet{bot97} claimed that 
the velocity dispersions inferred in the $B$-band trace the old stellar population because
a young population has far fewer absorption lines and they are located at the blue part
of the optical spectrum \citep[see also][]{bot93}.  However, a contamination by young stars with 
low velocity dispersion
would lead to an underestimate of the velocity dispersion. In fact, \citet{fuc99} estimated
that in the solar neighbourhood, the velocity dispersion derived in the $B$-band is
about $25\%$ smaller than the mass-weighted velocity dispersion. For bluer galaxies,
the $B$-band velocity dispersion could underestimate the mass-weighted velocity dispersion
by a larger fraction. If galaxies were observed in the near $V$-band,
as those used by \citet{ber11}, K-giants dominate
the velocity dispersion signal \citep{ani16}. In the solar neigbourhood, an
external observer that would fit the vertical velocity distribution with a single
Gaussian would obtain a vertical velocity dispersion of $13.0\pm 0.1$ km s$^{-1}$, which
would translate to $\sigma_{R}=22$ km s$^{-1}$ \citep{ani16}, which is $50\%$ smaller 
than the radial velocity dispersion of the old disc stars ($\simeq 44$ km s$^{-1}$, Binney
\& Tremaine 1987; Fuchs 1999).

It is worthwhile noting that, under MOND, a vertical velocity dispersion higher 
(by $30\%$ on average) than derived from population-integrated line profiles, 
is also required to explain the thickness of $30$ disc galaxies from the DiskMass survey 
\citep{ang15,mil15}.

\section{Discussion and conclusions}
\label{sec:conclusions}

We have analysed the dynamics of six late-type spiral galaxies
in standard dark matter models and under MOND. The selected galaxies
are bulgeless, show no evidence for strong bars, and they have high-quality measurements
of the rotation curves, derived from two-dimensional velocity fields.
In the sample, we have galaxies with relatively large internal
accelerations ($\ga a_{0}$) but cataloged as dark matter dominated
galaxies (S05 and \S \ref{sec:RCs_DM}).
Our main objective was to explore the capability of dark matter models
and MOND to explain simultaneously the shape of the rotation curves and the level 
of gravitational stability of the discs with realistic stellar mass-to-light ratios. 
In particular, we examined the bar stability of these galaxies using the ENL and the
PWC criteria. We summarize our main results below.

i)  We found that for all the galaxies in our sample, one can adjust 
the dark halo parameters and the stellar mass-to-light ratio of the disc 
to match the rotation curves and the stability requirements.

ii) In the dark matter models, the bar stability criteria place upper
limits on the value of $\Upsilon_{\star}$. We find that the stellar
discs in two galaxies of our sample (NGC 3621 and NGC 6503) are sub-maximal, 
which means that the disc is not as massive as allowed by the rotation curve.
In the case of NGC 6503, the upper limits derived with either ENL or PWC
criteria are consistent with previous work.

iii) We have compared the upper limits on $\Upsilon_{\star}$
derived from stability arguments with population-synthesis values. 
We find that there are two galaxies, NGC 3621 and NGC 5949, 
where the upper $\Upsilon_{\star}$-value derived from stability analysis is
a factor of $\sim 1.5-2$ lower than the population-synthesis value. 
Given that population-synthesis models have other sources of uncertainties
besides the IMF \citep[e.g.,][]{con13},
we conclude that dark matter models are consistent with observations for all
the galaxies in our sample.

iv) There is some tension between the observed rotation curve and the MOND fit
in three galaxies (NGC 4605, 5963 and 6503). These discrepancies cannot be
explained by reasonable changes in the adopted value of $a_{0}$ or in the 
distances. For NGC 4605 and NGC 5963,
MOND underpredicts the rotational speed at the outer parts of the observed rotation
curve. 
In particular, in the case of NGC 5963, MOND predicts a gently declining
rotation curve in the outer parts because the baryonic distribution is rather
compact and the internal accelerations are larger than $a_{0}$. However, 
observations suggest that the rotation curve is slowly rising. It would be
worthwhile exploring whether the level of non-circular motions in these galaxies can
affect the measured rotation curves.

v) In MOND, bar-stability requirements place lower limits to the stellar velocity dispersion.
We find that the stellar discs in these galaxies should be dynamically 
hot, which may have observable consequences. In most of the galaxies (5 out of 6 galaxies), 
the required radial velocity dispersion of the stellar disc to be bar stable is significantly 
larger than the mean value observed in other galaxies with similar rotation velocity.

vi) Since the selected galaxies have central surface brightness as normal spiral galaxies,
measurements of the stellar velocity dispersion until one radial scalelength 
should be feasible with mid-size telescopes.

vii) For one galaxy (NGC 6503), there is measurements of the stellar velocity dispersion.
Taking the data at face value, the observed velocity dispersion is inconsistent with
the value required to be bar stable in MOND.

viii) An alternative could be that the currently-available stellar velocity dispersions in external
galaxies are not representative of the bulk of the stellar mass. This should be
investigated further because it might have implications not only for MOND but also
for Newtonian dark matter models. 

Here we have focused on the global stability of the stellar disc in
galaxies where the contribution of the gas to the mass budget is
small. Low-mass gas-rich galaxies may be also interesting targets to
check stability. S\'anchez-Salcedo et al. (1999, 2014) found that
gas-rich dwarf galaxies as IC 2574, NGC 1560 and Holmberg II, with measured
H\,{\sc i} velocity dispersions, are marginally stable in MOND and
thus very responsive to perturbations.
It remains a challenge for MOND to explain why
these gas-rich galaxies have maintained a low rate of star formation.

\section*{Acknowledgements}
We thank Carlo Nipoti for thoughtful comments and insightful suggestions on 
the manuscript. This work was partly supported by CONACyT project CB-165584.
We acknowledge the usage of the HyperLeda database (http://leda.univ-lyon1.fr).
The authors made use of the NASA/IPAC Extragalactic Database (NED) which is operated by the Jet Propulsion Laboratory, California Institute of Technology, under contract with the National Aeronautics and Space Administration, THINGS `The H\,{\sc i} Nearby Galaxy Survey'
(Walter et al. 2008) and SDSS Data DR7. 
Funding for the SDSS and SDSS-II has been provided by the Alfred P. Sloan Foundation, the 
Participating Institutions, the National Science Foundation, the U.S. Department of Energy, the National 
Aeronautics and Space Administration, the Japanese Monbukagakusho, the Max Planck Society, and the 
Higher Education Funding Council for England. The SDSS Web Site is http://www.sdss.org/.
This research has also made use of the NASA/IPAC Infrared Science Archive, which is operated by the Jet Propulsion Laboratory, California Institute of Technology, under contract with the National Aeronautics and Space Administration.

{}

\appendix

\section{The MOND $X_{m}$-parameter in terms of Newtonian variables: the fictious dark halo}
\label{sec:app_X2_Newton_vars}
Consider a purely stellar disc in MOND. The stability parameters $X_{m}$
depend on the rotation curve and its derivatives through $\kappa$, because
conservation of angular momentum and differential rotation oppose to 
radial compressions in the disc. To go further, it is convenient to
write $\kappa^{2}$ in terms of $\kappa_{N}^{2}$, which is defined as
\begin{equation}
\kappa_{N}^{2}= R\frac{d\Omega^{2}_{N}}{dR}+4\Omega_{N}^{2},
\end{equation}
where $\Omega_{N}(R)\equiv v_{c,N}/R$. Thus, $\kappa_{N}$ is the epicyclic frequency of
the corresponding bare Newtonian disc (without the contribution of a dark halo).
Using the identity $\Omega^2=\Omega^{2}_{N}/\mu$,
we obtain
\begin{equation}
\frac{d\Omega^{2}}{dR}=\frac{1}{\mu}\frac{d\Omega_{N}^{2}}{dR}-\frac{\Omega_{N}^{2}}{\mu^{2}}
\frac{d\mu}{dR},
\end{equation}
and
\begin{equation}
\kappa^{2}=\frac{1}{\mu} \left(\kappa_{N}^{2}-\frac{\gamma_{N}L}{1+L}\Omega^{2}_{N}\right),
\label{eq:kappaMN}
\end{equation}
where 
\begin{equation}
\gamma_{N}\equiv \frac{d\ln g_{N}}{d\ln R}.
\end{equation}
In the derivation of Equation (\ref{eq:kappaMN}), we used the following identity: $\gamma_{N}=(1+L)\gamma_{M}$,
where $\gamma_{M}$ is the radial logarithmic derivative of $g_{M}$.
Substituting Equation (\ref{eq:kappaMN}) into Equation (\ref{eq:XmMOND}), we can write $X_{m}$ as
\begin{equation}
X_{m}=\frac{\mu^{+}}{\mu}(1+L^{+})^{1/2} \left(1-\frac{L}{1+L}
\left[\frac{\gamma_{N}}{3+\gamma_{N}}\right]\right)
\frac{R\kappa_{N}^{2}}{2\pi G m\Sigma}.
\end{equation}
For a disc with $\mu^{+}\simeq \mu$ and $L^{+}\simeq L$, this expression can be
simplified to
\begin{equation}
X_{m}\simeq \frac{1}{(1+L)^{1/2}} \left(1+\frac{3L}{3+\gamma_{N}}\right)
\frac{R\kappa_{N}^{2}}{2\pi G m\Sigma}.
\end{equation}
Thus, a galaxy under MOND has the same $X_{m}$-value than a Newtonian disc embedded
in an inert dark halo with a mass fraction $f_{d}(R)$
\begin{equation}
f_{d}\equiv \frac{M_{d}(R)}{M_{t}(R)}\simeq (1+L)^{1/2}\left(1+\frac{3L}{3+\gamma_{N}}\right)^{-1},
\end{equation}
where $M_{d}(R)$ and $M_{t}(R)$ are, respectively, the mass of the disc and the total mass interior to $R$.

The smallest value of $f_{d}$ occurs for galaxies in the deep MOND limit ($L\rightarrow 1$) and
in regions where the Newtonian rotation curve of the disc decays in a Keplerian fashion ($\gamma_{N}=-2$).
In that situation, $f_{d}\simeq 0.35$, which implies that the $X_{m}$-parameter in MOND is similar
to add a dark halo $2$ times as massive as the disc. 

For galaxies with $x\simeq 1$ (hence $L\simeq 0.5$) and 
at intermediate radii, say $R\simeq 2.2h_{R}$ (so that $\gamma_{N}\simeq -1$), we have $f_{d}=0.7$.
This means that the value
of $X_{m}$ is similar to that in a Newtonian disc when it is immersed in a dark halo with a mass of only 
$\sim 0.4M_{d}$ within $2.2h_{R}$.

\section{The stability parameters in QUMOND}
\label{sec:X2QUMOND}
In the quasi-linear formulation of MOND (QUMOND, Milgrom 2010), the field equation is given by
\begin{equation}
\nabla^{2} \Phi = \na \cdot \left[\nu (|\na \Phi_{N}|/a_{0}) \na \Phi_{N}\right),
\label{eq:PoissonQUMOND}
\end{equation}
where $\Phi_{N}$ is the Newtonian gravitational potential, which satisfies the
Poisson equation
\begin{equation}
\nabla^{2}\Phi_{N}=4\pi G\rho.
\end{equation}
Here, $\nu (y)$ satisfies that $\nu\rightarrow 1$ when $y\gg 1$ (Newtonian limit)
and $\nu\approx y^{-1/2}$ when $y\ll 1$.

In this Appendix, we are going to derive the change in the
gravitational potential of a disc when we make a linear perturbation
in the surface density. The procedure is standard and is frequently
used to calculate the dispersion relation in discs (e.g., Binney \& Tremaine 1987; 
Milgrom 1989; Roshan \& Abbassi 2015). It is assumed that the disc is infinitesimally thin
and use the tight winding approximation (WKB approximation).

The gravitational potential and forces in the disc are separated into the unperturbed part 
(denoted by a subscript $0$) and the perturbations:
\begin{equation}
\Phi=\Phi_{0}+\phi,
\end{equation}
\begin{equation}
\Phi_{N}=\Phi_{N,0}+\phi_{N},
\end{equation}
\begin{equation}
\vecg_{M}=\vecg_{M,0}+\vecq_{M},
\end{equation}
and
\begin{equation}
\vecg_{N}=\vecg_{N,0}+\vecq_{N}.
\end{equation}
Assuming that the perturbations are small, we linearize the QUMOND field equation 
(\ref{eq:PoissonQUMOND}) and obtain
\begin{equation}
-\nabla^{2}\phi=-\nu_{0}\nabla^{2}\phi_{N}+\vecq_{N}\cdot\na \nu_{0}+\na\left[\tilde{\nu}_{0} (\vecq_{N}\cdot \hat{\vece}_{0}) \hat{\vece}_{0}\right],
\label{eq:linear_complete}
\end{equation}
where $\tilde{\nu}\equiv y\nu'$ and $\hat{\vece}_{0}\equiv \vecg_{N,0}/g_{N,0}$ is the unit 
vector parallel to $\vecg_{N,0}$. Initially, we assume that the disc is axisymmetric. In that
case,  $\hat{\vece}_{0}$ has only radial ($e_{R}$) and vertical ($e_{z}$) components.

We wish to calculate $\phi$ when the the perturbed surface density $\Sigma_{1}$ has the form of a tightly 
wound wave, that is
\begin{equation}
\Sigma_{1}=\Sigma_{\alpha}\exp(ikR-i\omega t), 
\end{equation}
with $|kR|\gg 1$. This WKB approximation
permits us to simplify calculations by neglecting terms proportional to $1/R$ compared with 
the terms proportional to $k$; thus the wave resembles locally a plane wave. 
As the perturbed quantities vary more quickly than the unperturbed
variables, the gradients of the perturbed quantities are larger and so that we
only retain derivatives of the perturbed quantities in Equation (\ref{eq:linear_complete}). 
Outside the disc, $\nabla^{2}\phi_{N}=0$, and the resultant differential equation for
$\phi$ is 
\begin{equation}
\nabla^{2}\phi=
\tilde{\nu}_{0}\left[(e_{z}^{2}-e_{R}^{2})\frac{\partial^{2}\phi_{N}}{\partial z^{2}}
+2e_{z}e_{R}\frac{\partial^{2}\phi_{N}}{\partial R\partial z}\right].
\label{eq:nablaWKB}
\end{equation}

From the standard Newtonian analysis (e.g., Binney \& Tremaine 1987), the perturbation
on the Newtonian potential is given by
\begin{equation}
\phi_{N}=-\frac{2\pi G}{|k|}\Sigma_{\alpha}\exp(ikR-|kz|-i\omega t).
\end{equation}
Once we know $\phi_{N}$, we can compute its spatial derivatives and after substituting
into Equation (\ref{eq:nablaWKB}), we derive a differential equation that obeys $\phi$
outside the disc:
\begin{equation}
\nabla^{2}\phi=2\pi G\tilde{\nu}_{0}{\mathcal{E}}_{0}
|k|\Sigma_{\alpha}\exp(ikR-|kz|-i\omega t),
\label{eq:diff_eq}
\end{equation}
where 
\begin{equation}
{\mathcal{E}}_{0}=1-2e_{z}^{2}-2{\rm sgn}(k)ie_{R}|e_{z}|,
\end{equation}
and ${\rm sgn}$ is the sign function.
 The solution of Equation (\ref{eq:diff_eq}) that decays at large heights from the disc is given by 
\begin{eqnarray}
&&\phi=\phi_{\alpha}^{(1)}\exp(ikR-|kz|-i\omega t)\\ \nonumber
&&+2\pi G \tilde{\nu}_{0}{\mathcal{E}}_{0}\frac{|k|}{k_{0}^{2}-k^{2}}
\Sigma_{\alpha}\exp(ikR-|k_{0}z|-i\omega t),
\label{eqn:solution_coef}
\end{eqnarray}
where $\phi_{\alpha}^{(1)}$ and $k_{0}$ are determined in
the following by imposing the boundary conditions at $z=0$.

Integrating Equation (\ref{eq:linear_complete}) in a short cylinder whose flat faces are parallel
to the disc and located at $z=\pm h$, with $h\rightarrow 0$, we find that 
\begin{equation}
\frac{\partial \phi}{\partial z}\bigg|_{z=0^{+}}=\nu_{0}^{+}\frac{\partial \phi_{N}}{\partial z}\bigg|_{0^{+}}+\tilde{\nu}_{0}^{+}
\left(\frac{\partial \phi_{N}}{\partial z}\bigg|_{0^{+}} e_{z}^{+}+\frac{\partial \phi_{N}}{\partial R}
\bigg|_{0^{+}}e_{R}\right)e_{z}^{+},
\label{eq:boundary}
\end{equation}
where $\nu_{0}^{+}$, $\tilde{\nu}_{0}^{+}$ and $e_{z}^{+}$ are the values of $\nu_{0}$,
$\tilde{\nu}_{0}$ and $e_{z}$ just above the disc.
Substituting the $R-$ and $z-$derivatives of $\phi_{N}$ into Equation (\ref{eq:boundary}), we get
\begin{eqnarray}
&&\frac{\partial \phi}{\partial z}\bigg|_{z=0^{+}}=2\pi G\Sigma_{\alpha}
\exp(i[kR-\omega t])\\ \nonumber
&&[\nu_{0}^{+}+\tilde{\nu}_{0}^{+}
(e_{z}^{+})^{2}+\tilde{\nu}_{0}^{+} {\rm sgn} (k) i e_{R}|e_{z}^{+}|].
\label{eqn:QU_phi_z2}
\end{eqnarray}
By equating Equation (\ref{eqn:QU_phi_z2}) with the $z$-derivative of Equation (\ref{eqn:solution_coef}),
we obtain 
\begin{equation}
\frac{|kk_{0}|}{k_{0}^{2}-k^{2}}=\frac{1}{2},
\end{equation}
which implies that $k_{0}=\pm(1+\sqrt{2})k$ (note that the choice of the sign of $k_{0}$ is
physically irrelevant). We also obtain that
\begin{equation}
-|k|\phi_{\alpha}^{(1)}-\pi G \tilde{\nu}_{0}^{+}\Sigma_{\alpha}=2\pi \nu_{0}^{+}G\Sigma_{\alpha},
\end{equation}
implying
\begin{equation}
\phi_{\alpha}^{(1)}=-\frac{2\pi (\nu_{0}^{+}+\tilde{\nu}_{0}^{+}/2)G\Sigma_{\alpha}}{|k|}.
\label{eq:phialpha1}
\end{equation}

Combining Eqs. (\ref{eqn:solution_coef}) and (\ref{eq:phialpha1}) and using that 
$k_{0}=\pm(1+\sqrt{2})k$, we finally find the perturbation on
the gravitational potential in the plane of the disc:
\begin{equation}
\phi(R,0)=-\frac{2\pi \chi_{\scriptscriptstyle Q} G\Sigma_{1}}{|k|}, %G \Sigma_{\alpha}\exp(ikR-i\omega t)
\end{equation} 
with
\begin{equation}
\chi_{\scriptscriptstyle Q}=\nu_{0}^{+}+\frac{\tilde{\nu}_{0}^{+}}{2}  \left(1-\frac{{\mathcal{E}}_{0}^{+}}{1+\sqrt{2}}\right).
\end{equation}
Therefore, the linear perturbation on the QUMOND potential, the QUMOND Toomre criteria and 
the $X_{m}$-parameters can be obtained by replacing $G$ for
$\chi_{\scriptscriptstyle Q} G$ in the respective Newtonian formulae
\begin{equation}
Q=\frac{\sigma_{R}\kappa}{\pi \chi_{\scriptscriptstyle Q}G\Sigma},
\end{equation}
\begin{equation}
X_{m}=\frac{R\kappa^{2}}{2\pi m \chi_{\scriptscriptstyle Q}G\Sigma}.
\end{equation}

In a thin rotating disc, the $z$-component of the acceleration is much smaller than the local
radial acceleration and hence ${\mathcal{E}}_{0}^{+}=1$. In that case
\begin{equation}
\chi_{\scriptscriptstyle Q}=\nu_{0}^{+}+\left(1-\frac{1}{\sqrt{2}}\right)\tilde{\nu}_{0}^{+}=
\nu_{0}^{+}\left(1+\left[1-\frac{1}{\sqrt{2}}\right]{\mathcal{L}}_{0}^{+}\right),
\end{equation}
where ${\mathcal{L}}=y\nu'/\nu$.

In the Newtonian limit, $\nu_{0}^{+}=1$ and ${\mathcal{L}}_{0}=0$. Consequently $\chi=1$ 
and the classical Newtonian case is recovered. In the deep MOND regime, ${\mathcal{L}}_{0}=-1/2$,
and then
\begin{equation}
\chi_{\scriptscriptstyle Q}=\frac{\nu_{0}^{+}}{2}\left(1+\frac{1}{\sqrt{2}}\right).
\end{equation}

In order to compare QUMOND with the non-linear Poisson formulation regarding the degree of stability,
it is convenient to write down $\chi_{\scriptscriptstyle Q}$ in terms of $\mu^{+}$ and $L^{+}$
using the relations
\begin{equation}
\nu(y) =\frac{1}{\mu(x)},
\end{equation}
and
\begin{equation}
{\mathcal{L}}=-\frac{L}{1+L}.
\end{equation}
We obtain
\begin{equation}
\frac{\chi_{{\scriptscriptstyle Q}}}{\chi_{nl}}=\frac{1}{(1+L^{+})^{1/2}}\left(1+\frac{L^{+}}{\sqrt{2}}\right).
\end{equation}
It holds that $\chi_{\scriptscriptstyle Q}\geq \chi_{nl}$, suggesting that self-gravitating discs are more
stable in the non-linear Poisson formulation than they are in QUMOND. For discs with $L^{+}\simeq 0.5$,
as those considered in this work, $\chi_{\scriptscriptstyle Q}/\chi_{nl}\simeq 1.1$.

\section{Model inference using a Bayesian approach}
\label{sec:statistics}
Bayesian inference relies on the concept of conditional probability to
revise one's knowledge. Prior to the collection of sample data one had some
(perhaps vague) information on the parameter of interest $\theta$. This
initial uncertainty is formally modelled via a \textit{prior distribution}
for the parameter $\theta$. In our case the parameter of interest is 
$\Upsilon _{\star }$. Then using Bayes Theorem we can combine the model
density of the observed data with the prior density to get the \textit{
posterior density}, that is, the conditional density of $\theta $ given the
data. Until further data is available, this posterior distribution of 
$\theta$ is the only relevant information as far as is concerned.

Some notation is required to describe the Bayesian perspective. Let us
denote the prior distribution by $\pi (\theta )$. Let $n$ be the number of
observations for a given galaxy. The model for the observed data 
$\mathbf{y}=({y_{1},y_{2},\ldots ,y_{n})}$ is expressed by the likelihood function 
$\mathcal{L}(\theta )=f(\mathbf{y}|\theta )$. The joint distribution of the
data and the parameter of interest is given by:
\begin{equation*}
f(\mathbf{y},\theta )=f(\mathbf{y}|\theta )\pi (\theta )=\mathcal{L}(\theta
)\pi (\theta ).
\end{equation*}

One can update the knowledge on $\theta $ by computing the conditional
distribution of $\theta $\ given the information in the sample $\mathbf{y}$.
This is done via Bayes Theorem that states that the {\emph{posterior
distribution}} of $\theta $ is given by
\begin{equation}
\pi (\theta |\mathbf{y})=\frac{f(\mathbf{y}|\theta )\pi (\theta )}{\int f(
\mathbf{y}|\theta )\pi (\theta )d\theta }.  \label{eq:posterior}
\end{equation}

There is\ a simulation method called Markov Chain Monte Carlo that generates
a sequence of values of $\theta ,$\ say $\left\{ \theta _{i}\right\}_{i=1}^{MC},$\ 
that has a distribution equal to (\ref{eq:posterior}). This
method has the advantage that avoids the need to compute the integral in the
denominator of the right-hand-side of Equation (\ref{eq:posterior}). The method just requires the
specification of the likelihood function and the prior density. The value of 
$MC$\ is usually huge, say $100,000$.

For this Bayesian approach we assume the simplest prior distribution for the
unknown parameter $\theta$, that is, the Jeffrey's prior (also known as
uniform or flat prior). The upper limit for this distribution was obtained
using stability arguments to provide upper limits on the surface density of
stellar disks, as mentioned in the introduction. For the likelihood function
we assume that the discrepancies between the rotation velocities and the
model in question have a normal distribution. This can be expressed as
\begin{equation*}
f(\mathbf{y}|\theta )\propto \exp \biggl({-\frac{1}{2}\sum_{i=1}^{n}
\biggl[\frac{y_{i}-\mu (x_{i},\theta )}{\sigma _{i}}\biggr]^{2}}\biggr),
\end{equation*}
where $y_{i}$ is the observed rotation velocity at the corresponding radius 
$x_{i}$, $\mu (x_{i},\theta )$ is a nonlinear specification that depends on
the astro-dynamics of each galaxy and $\sigma _{i}^{2}$ is the measurement
error.

Once we obtain a sequence of values $\left\{ \theta _{i}\right\} _{i=1}^{MC}$
from the posterior distribution $\pi (\theta |\mathbf{y})$\ we use them to
make inferences on a nonlinear, non-monotonic transformation $g(\theta )$ by
computing the sequence $\left\{ g(\theta _{i})\right\} _{i=1}^{MC}$. From
this we calculate a point estimator using the average, 
$\frac{1}{MC}\sum_{i=1}^{MC}g(\theta _{i})$, or the median. In order to obtain a 
$(1-\alpha )\%$ interval that reflects the uncertainty in the estimation we
used the percentiles of $\frac{\alpha }{2}\%$ and $(1-\frac{\alpha }{2})\%$
of this sequence to obtain a posterior credibility interval for $g(\theta)$.

For example, $X_{2}(R,\Upsilon _{\star})$\ is a highly nonlinear function
of $\Upsilon _{\star}$. Using the above method we were able to obtain the
bands in Figure \ref{fig:X2_MOND}. Also $\left\langle X_{2}\right\rangle$\ is the average
of a highly nonlinear function of $\Upsilon _{\star}$. Using the above
method we were able to obtain posterior credibility intervals for its value,
which are given in Table \ref{table:MONDvalues}.

It is important to stress the relevance of using the Bayesian approach for
the models studied in this paper. The models under consideration are highly non-linear
functions of the parameter of interest. In statistics this is usually
treated using approximate methods that assume a large sample, that is a
value of $n$ in the order of hundreds. This is not the case in our samples;
hence those approximations are expected to produce inconsistent results in
these models. For example, using the large sample methods, the bands for 
$X_{2}(R,\Upsilon _{\star})$\ or the interval for $\left\langle X_{2}\right\rangle$\ 
could very well include negative values.

\end{document}